\documentstyle[11pt,aaspp4]{article}

\def\etal{{\it et al.}}
\def\x{$\xi(r)$}
\def\xip{\ifmmode\xi(r_p,\pi)\else $\xi(r_p,\pi)$\fi}
\def\wp{$w_p(r_p)$}
\def\kms{\ifmmode\,{\rm km}\,{\rm s}^{-1}\else km$\,$s$^{-1}$\fi}
\def\hmpc{\ifmmode\,h^{-1}\,{\rm Mpc}\else$h^{-1}\,$Mpc\fi}
\def\iras{{\sl IRAS\/}}
\def\sig{$\sigma_{12}(r)$}
\def\sig1{$\sigma_{12}(1)$}
\def\bfv{{\bf v}}
\def\ro{r_0}

\def\ltsima{$\; \buildrel < \over \sim \;$}
\def\simlt{\lower.5ex\hbox{\ltsima}}
\def\gtsima{$\; \buildrel > \over \sim \;$}
\def\simgt{\lower.5ex\hbox{\gtsima}}

\lefthead{Guzzo et al.}
\righthead{Redshift Distortions}

\begin{document}

\title{Redshift--Space Distortions and the Real--Space Clustering\\
of Different Galaxy Types}

\author{Luigi Guzzo\altaffilmark{1,2}, Michael A. Strauss\altaffilmark{1,3},
Karl B. Fisher\altaffilmark{4,5},}
\author{Riccardo Giovanelli\altaffilmark{6}, and 
Martha P. Haynes\altaffilmark{6}}


\altaffiltext{1}{Department of Astrophysical Sciences, Princeton 
University, Princeton, NJ 08544.}
\altaffiltext{2}{Permanent Address: Osservatorio Astronomico di Brera, 
Via Bianchi 46, I--22055 Merate (LC), Italy.} 
\altaffiltext{3}{Alfred P. Sloan Foundation Fellow.}
\altaffiltext{4}{Institute for Advanced Study, Olden Lane, Princeton,
NJ 08540}
\altaffiltext{5}{Current Address: Signal Physics Group, Applied
Research Laboratories, The University of Texas, Austin, Texas, 78713} 
\altaffiltext{6}{Department of Astronomy and Center of Radiophysics and
Space Research, Space Sciences Bldg., Cornell University, Ithaca, NY 14853}


\begin{abstract}
We study the distortions induced by peculiar velocities on the 
redshift--space correlation function of 
galaxies of different morphological
types in the Pisces--Perseus redshift survey.  Redshift--space
distortions affect early-- and late--type galaxies 
in different ways.  In particular, at small separations, the dominant
effect comes from virialized cluster cores, where ellipticals are the 
dominant
population.  The net result is that a meaningful comparison of the clustering
strength of different morphological types can be performed only in real space,
i.e., after projecting out the redshift distortions on the two--point 
correlation function \xip.  A power--law fit to the projected function \wp\
on scales smaller than 10 \hmpc\ gives $\ro = 8.35_{-0.76}^{+0.75}$ \hmpc, $\gamma = 2.05_{-0.08}^{+0.10}$ for
the early--type population, and $\ro = 5.55_{-0.45}^{+0.40}$ \hmpc, $\gamma = 
1.73_{-0.08}^{+0.07}$ for spirals and irregulars.
These values are derived for a sample luminosity limited to $M_{Zw}\le
-19.5$.  We detect a 25\% increase of $\ro$ with luminosity for all
types combined, from
$M_{Zw} = -19$ to $-20$.  In the framework of a simple stable--clustering model
for the mean streaming of pairs, we estimate \sig1, the one--dimensional 
pairwise velocity dispersion between 0 and $1 \hmpc$,
to be $865^{+250}_{-165} \kms$ 
for early--type galaxies and $345^{+95}_{-65}\kms $
for late types.   This latter value should be a fair
estimate of the pairwise dispersion for ``field'' galaxies; it is stable 
with respect to
the presence or absence of clusters in the sample, and is 
consistent with
the values found for non--cluster galaxies and \iras\ galaxies at similar
separations.

\end{abstract}


\keywords{}


%

\section{INTRODUCTION}

Peculiar velocities distort the maps of the galaxy distribution
when redshifts are used as a measure of 
distance through the Hubble relation. The observed distortions 
contain important information on the statistical 
properties of the large--scale motions of galaxies, presumably due to
the gravitational influence of the
true underlying mass distribution.  In particular, the two--point correlation
function in redshift space $\xi(s)$ differs from that
in real space \x\ in two respects: on small scales correlations
are suppressed due to the virialized motions in rich clusters,
which in redshift space elongate structures along the line of sight;
on large scales coherent motions produced by infall into
overdense regions or by outflow out of underdense regions 
enhance correlations.   

Galaxies of different morphological types inhabit different environments,
following a well--established morphology--density relation (e.g., Dressler 
1980; Postman \& Geller 1984).
As a consequence, they display significantly different clustering properties 
(Davis \& Geller 1976; Giovanelli, Haynes \& Chincarini 1986; Iovino \etal\
1993; Loveday \etal\ 1995; Hermit \etal\ 1996): ellipticals and S0's dominate
dense cluster cores and are therefore more clustered than spirals and 
irregulars.  However, their association with the deep potential wells
of clusters implies that they have higher peculiar
velocities on average, so that the small--scale $\xi(s)$ for early-type 
galaxies is more strongly suppressed with respect
to \x\ than for late types.  The consequence is that a correct comparison 
of the clustering properties of different morphological types requires 
understanding in detail their respective redshift space distortions.
One way to avoid this problem is to measure the angular correlation
function $w(\theta)$ on two--dimensional catalogues (Giovanelli \etal\
1986; Loveday \etal\ 1995\footnote{These latter authors
perform also a cross--correlation between morphological subsamples of
the APM--Stromlo redshift survey and the APM angular catalogue, see 
\S~\ref{sec:xi(r)}.}).  The effects of redshift space distortions also
need to be quantified in comparing the angular clustering of
distant objects with the clustering in redshift space of galaxies at
low redshift (e.g., Iovino \etal\ 1996). 

The standard way to quantify redshift distortions is to split the separation 
vector of a pair of objects into components lying on the plane 
of the sky, $r_p$, and along the line of sight, $\pi$, and to compute 
the correlation
function \xip\ as a function of these two components.
The iso--correlation contours of \xip\ will be stretched along the 
$\pi$ direction at small separations, due to the effect of large 
velocity dispersions, and compressed at large scales, as a consequence
of large-scale coherent motions.  
Projecting \xip\ onto the $r_p$ axis gives the projected function
$w_p(r_p)$, which is independent of redshift distortions and can be 
directly expressed as an integral over the real--space correlation 
function $\xi(r)$.  The equation relating $w_p(r_p)$ and $\xi(r)$ thus
allows one to recover the latter via direct inversion or modelling. 

On the other hand, modelling the distortions of \xip\ allows
one to characterize the pairwise velocity distribution function.
We are particularly concerned with the second 
moment of this distribution function, $\sigma_{12}(r)$, in this paper. 
Davis \& Peebles (1983, hereafter \cite{dp83}) used the CfA1 survey
(\cite{Huchra83}) to measure \sig1, the value of $\sigma_{12}(r)$ at 
$r = 1\,\hmpc$, to be $340 \pm 40\kms$.  
While the analysis of the \iras\ 1.2 Jy redshift survey by Fisher
\etal\ (1994b, hereafter \cite{f94b})
produced a similar result, $\sigma_{12}(1) =317^{+40}_{-49} \kms$, 
re--analyses of the CfA1 survey (\cite{mo,zur,som}), and of larger optical
redshift surveys have shown a large range of values, going as high as
$1000 \kms$
(\cite{marzke}, \cite{g96}, \cite{lin}).  In particular, $\sigma_{12}(1)$
is found to be quite sensitive to the presence or absence of one or
two rich clusters, even in volumes as large as the CfA2. 
It seems plausible that
while the CfA1 value was strongly affected by the smallness of the
volume surveyed, that measured from the \iras\ 1.2 Jy survey reflects 
the specific nature of \iras\ galaxies, which are mostly star-forming,
late-type galaxies which are under-represented in rich clusters
relative to optically selected galaxies (\cite{Strauss92}). 

The case of \iras\ galaxies explicitly illustrates the dangers of
using a specific class of objects to draw conclusions on statistics of
the velocity field, in particular at small separations: the answer
depends sensitively on the morphological type of the tracer we are
using.  This paper addresses this issue in detail:  1) What
is the difference in the clustering strength of early-- and 
late--type galaxies measured in real space? 2) What is the 
small--scale velocity dispersion for the two classes of objects?

The outline of the paper is as follows: in \S~\ref{sec:data}, we
present the data we will use for our analyses.  We discuss the
measurement of \xip\ in \S~\ref{sec:estimating-xi}, and present our
results in \S~\ref{sec:xip}.  Our conclusions are summarized in
\S~\ref{sec:summary}.

\begin{table*}
\begin{center}
\begin{tabular}{llrcr}  \hline\hline 
Sample  & $M_{lim}$ & $d_{lim}(\hmpc)$ & Morphology & $N_{gal}$ \\ 
\hline
PP--19   	& --19     & 79  & All & 1021 \\
PP--19.5 	& --19.5   & 100 & All & 852 \\
PP--19.5--NOP 	& --19.5   & 100 (No Perseus) & All & 803 \\
PP--20   	& --20     & 126 & All & 577  \\
S--19    	& --19     & 79  & Late& 565  \\
S--19.5  	& --19.5   & 100 & Late& 481  \\
S--19.5--NOP  	& --19.5   & 100 (No Perseus) & Late& 458  \\
S--20    	& --20     & 126 & Late& 333  \\
E--19.5  	& --19.5   & 100 & Early& 278  \\
\hline
\end{tabular}
\end{center}
\caption{Properties of the volume--limited subsamples analyzed.}
\label{tab-samples}
\end{table*}                                                                   

\section{THE DATA: DEFINITION OF THE SAMPLES}
\label{sec:data}

We use the Perseus--Pisces redshift catalogue (cf., \cite{GH91}),
which includes redshifts for all Zwicky galaxies (\cite{Zwickycat}) in the 
positive--declination part of the South Galactic Cap (i.e., about $21^h \le 
\alpha \le 5^h$, $0^\circ \le \delta \le 50^\circ$).   
As \cite{ghc86} make clear, the Perseus-Pisces redshift survey is
affected by Galactic extinction around the edges.  For statistical studies,
therefore, it has to be properly restricted.  The Zwicky catalogue is 
nominally complete to $m_{Zw} = 15.7$: we thus impose an extinction--corrected 
magnitude cut of 15.5, trim the survey to $22^h \le \alpha \le 4^h$, 
$0^\circ \le 
\delta \le 42^\circ$, and apply the additional cut indicated by the heavy
line of Figure~\ref{survey_map}.  This excludes nearly all regions
with absorption $A_B > 0.2$, as given by the extinction maps of
\cite{BH78}, while it leaves in the core of the Perseus cluster ($\alpha \sim
3.2^h,\ \delta \sim +41^\circ$) to allow us
to study the robustness of our results to the presence of the richest
cluster in the region.  The magnitude--limited sample selected in this
way contains 4111 galaxies.  One potential problem of our 
selection criteria
is that they push the Zwicky catalogue to its completeness limit. 
In particular, with the extinction correction we include galaxies 
with observed Zwicky magnitudes $m_{Zw} \simgt 15.5$, where magnitude errors 
are large (e.g., Bothun \& Cornell 1990).  We shall show 
in \S~\ref{sec:xi(r)} that our principal results are indeed quite robust
to these uncertainties in the parent photometric catalogue. 

%
%
All the analyses of this paper are done with volume-limited subsamples of the
data.  That is,
we select a lower--limiting luminosity (or 
equivalently, an upper limit in absolute magnitude), and a
corresponding maximum distance implied by our apparent magnitude
limit, giving us uniform sampling throughout the volume.  This has the
effect of de-emphasizing the Pisces-Perseus chain relative to the
magnitude-limited case, because in the latter, the selection function
peaks near the redshift of the supercluster.  Also, this choice
is crucial for discussing luminosity effects, and eliminates uncertainties
related to the weighting schemes necessary when analysing magnitude--limited
samples. 

Table~\ref{tab-samples} summarizes the parameters of the
volume-limited samples we have used.  The range of absolute
magnitudes covered by the subsamples reflects the need 
to maximize our volume, while 
keeping
a sufficient number of objects within it.  Absolute magnitudes 
were calculated assuming $H_0 =
100\, \kms\,\rm Mpc^{-1}$.  The E--19.5 sample (``ellipticals'') 
contains galaxies with early 
morphological types (E, S0, and S0a), while the S-- samples
(``spirals'') contain all galaxies classified as spirals
or irregulars.   The morphological information available in the
catalogue is in reality quite finer, subdivided into 14 classes. 
The morphological coding
is 
is from the UGC for those galaxies in that catalog, and has been
estimated from sky survey plates for the remainder of the galaxies 
(Giovanelli, Haynes \& Chincarini 1986). 
To maximize the 
statistics within the volume--limited samples, we restrict our analysis 
to the two broad groups of early-- and late--tape galaxies. 
The spiral class is however large enough to define several samples to
different absolute magnitudes (S--19, S--19.5, and S--20); we also
define a spiral sample trimmed to exclude the Perseus cluster (S--19.5--NOP). 
Finally, we define an equivalent sample including all morphologies for
comparison, PP--19.5--NOP.

The cone diagrams of Figure~\ref{cone_all} 
show the galaxy distribution in PP--19.5 and PP--20, in which 
all morphological types are included, while
Figure~\ref{cone_types} shows the
corresponding distribution of the E--19.5 and S--19.5 samples,
respectively.  Note that the Perseus-Pisces chain, the overdensity at
$cz \approx 6000 \kms$, is more prominent in the ellipticals than the
spirals, consistent with the observed distribution on the sky
(\cite{ghc86}).  This is made quantitative in 
Figure~\ref{four_hist}, which shows the redshift histogram of each of
these subsamples; the dashed line in each case, proportional to $r^2$,
gives the expected 
distribution in the absence of structure.


\section{ESTIMATING THE TWO--POINT CORRELATION FUNCTION IN REDSHIFT SPACE}
\label{sec:estimating-xi}

\subsection{Definitions of $r_p$ and $\pi$}
\label{sec:xip-def}

The effect of redshift--space distortions can be understood through 
the correlation function \xip, where the radial separation of pairs
is split into two components: $\pi$, parallel to the line of sight,
and $r_p$, perpendicular to it.  There are two definitions of these
quantities in the literature.  Given two galaxies at redshifts $v_1$
and $v_2$, separated by angle $\theta$, \cite{dp83} define:
\begin{equation}
\pi \equiv{1\over H_0}\,\left|v_1 - v_2\right|\,\,\,\,\,\,\,\,\,\,\,\,\,\,
r_p \equiv{1\over H_0}\,\left(v_1 + v_2\right)\,\tan{\theta \over 2}\, .
\label{eq:dp83-def} 
\end{equation}
Note that the quadrature sum of $r_p$ and $\pi$ is {\em not\/}
equal to the redshift space distance $s$ between the galaxies. 
Recognizing this, Fisher \etal\ (1994a, hereafter \cite{f94a}) use a
slightly different definition.  They define the line of sight vector 
${\bf l} \equiv (\bfv_1 + \bfv_2)/2$ and the redshift difference
vector ${\bf s} \equiv \bfv_1 - \bfv_2$, leading to the definitions
\begin{equation}
\pi \equiv {{{\bf s} \cdot {\bf l}}\over{H_0 |{\bf l}|}}\quad\quad
r_p^2 \equiv {{\bf s} \cdot {\bf s}\over H_0^2} - \pi^2\, .
\label{f94-def}
\end{equation}
If we recast the \cite{f94a} formulation in terms of $\theta$, we
find:
\begin{equation}
\pi = {1 \over H_0} |{ v_1} - { v_2}| + {\cal O}(\theta^2)\quad\quad
r_p \simeq {1\over H_0}\,\left({ v_1} + { v_2}\right)\,\tan{\left(\theta 
\over 2\right)}\;{4{ v_1} { v_2}\over ({ v_1} + { v_2})^2}\, ,
\end{equation}
which shows explicitly that the two definitions are {\it not\/} strictly
equivalent, even in the small--angle approximation.
For our analysis we use Eq.~(\ref{f94-def}), but checks using
Eq.~(\ref{eq:dp83-def}) show only negligible differences in our \xip\
maps.  We conclude 
therefore that one can make direct comparison of results obtained with the two 
different definitions. 

\subsection{Measuring \xip}
\label{sec:methods}

We estimate the quantity \xip\ using the method of \cite{dp83}.  A
catalog of $n_R=100,000$ uniformly distributed points with the same boundaries
as the real sample is prepared.  We count the number of pairs in
1~\hmpc\ bins 
of $r_p$ and $\pi$ among the $n_G$ galaxies [$N_{GG}(r_p,\pi)$], and
between the galaxies and the random sample [$N_{GR}(r_p,\pi)$].  Our
estimate of the correlation function is then

\begin{equation}
\xip = {N_{GG}(r_p,\pi)\over N_{GR}(r_p,\pi)}{2\,n_R \over 
n_G} - 1 \,\, .
\label{xi_est}
\end{equation}

Because our samples are volume-limited, each galaxy gets equal
weight, and thus we do not apply the statistical weights needed
when analyzing magnitude--limited catalogues.  For the most part, we
confine ourselves to 
scales less than $10 \hmpc$, and thus there is little benefit to using
the alternative estimator of Hamilton (1993). This is less sensitive
than is Eq.~(\ref{xi_est}) to uncertainties in the mean density, and
therefore is important when measuring the correlation function on
very large scales.

\subsection{Error Estimation and Maximum--Likelihood Fits} 
\label{sec:ML}

Following \cite{Ling86}, we use bootstrap resampling to compute
statistical errors on our estimates of quantities we derive 
{}from \xip.  As we will see in the next section, we carry out
quantitative analyses not on \xip\ directly, but rather two derived
quantities: $w_p$ (Eq.~\ref{wp}, the projection of \xip\ on the $r_p$
axis, which does not suffer redshift-space distortions), and $\xi(\pi)$
(Eq.~\ref{eq:xipi}, essentially a cut in \xip\ at constant $r_p$ to
measure redshift space distortions).  We thus
compute errors, and the 
covariance matrix, of the 1-D quantities $w_p$ and $\xi(\pi)$, rather
than of the 2-D \xip. 
For each of the samples of Table~\ref{tab-samples}, we create 100
bootstrap realizations, and compute $w_p$ and $\xi(\pi)$ for each.
Determination of the covariance matrix of errors is then
straightforward, following \cite{f94a}. 
\cite{f94a} have discussed the method in detail, showing that it gives
a good representation of true statistical errors for the correlation
function on scales below $\sim 10 \hmpc$, but it tends to overestimate
them on larger scales. 

This covariance matrix enables us to 
fit models for the real space
correlation function $\xi(r)$ and the redshift distortions to the
observed $w_p$ and $\xi(\pi)$ via $\chi^2$ (cf., \cite{f94b}). 
In practice, the effective number of degrees of freedom in the data is smaller
than the number of sampled values of $w_p$ and $\xi(\pi)$ (i.e., these
functions are oversampled), and thus
the covariance matrix is singular.  We therefore follow \cite{f94b} in using
singular value decomposition, which allows the calculation of the
matrix product in the $\chi^2$ function in a robust manner.


\section{REDSHIFT--SPACE DISTORTIONS: \xip}
\label{sec:xip}

\subsection{The Observed \xip}\label{xp-shape}

Figures~\ref{xipi_first}--\ref{xipi_morp} display the observed \xip\ 
both for the complete samples PP--19, PP--19.5, PP--20, and for the
morphological subsamples E--19.5 and S--19.5.  For PP--19, we show 
the original \xip,
while for all samples, we show a version smoothed by a 
$3\hmpc\times3\hmpc$ Gaussian to suppress the binning noise and to
bring out the global features of the maps.  All the statistical
analyses below are of course carried out on the unfiltered data.

The contours for PP--19 and PP--19.5 are enormously distorted at very 
small scales, a signature of a high pairwise dispersion $\sigma_{12}(r)$
at small $r$, as we shall quantify in \S~\ref{sec:sigma-12}.  
Most of this distortion is produced by pairs lying in clusters,
in particular in the smallest sample, PP--19, which is
dominated by half--dozen rich clusters along the Perseus--Pisces chain.
The small--scale elongations are substantially smaller in the PP--20
sample.  This sample has a volume of $7.4 \times 10^5\;h^{-3}\;\rm
Mpc^3$; this is quite a bit larger than the volume of PP--19.5, but it
does not include any more clusters, and therefore the cluster
contribution to $\sigma_{12}(1)$ is somewhat diluted. 
This volume is still a factor of 6 smaller than the volume at which Marzke
\etal\ (1995) show -- using both a COBE--normalized Cold Dark Matter model 
and a phenomenological model based on the observed distribution of Abell 
cluster velocity dispersions -- that $\sigma_{12}(1)$ stabilizes.  For
samples of volume comparable to those used here, they derive a typical
uncertainty on $\sigma_{12}(1)$ of $\sim 180 \kms$.  Although this
scatter refers to 
non-overlapping samples, 
it does give an explanation for the observed difference in $\sigma$
for PP--19 and PP--20. 

On large scales, the compression of \xip\ for PP--19
(in particular the \xip$=0.2$ and \xip$=0$ contours) is that expected
due to large-scale streaming (cf., \cite{f94b}). 
However, the Perseus-Pisces supercluster (cf., Figure~\ref{cone_all})
lies largely in the plane of the sky, so that the real--space 
distribution of galaxies is intrinsically anisotropic in this sample.
Thus it is not a fair sample for measuring large-scale
streaming.  Some part of the observed distortions of the \xip\ 
contours must be due to infall onto the 
supercluster itself, as directly observed by Giovanelli \etal\ (1996), 
who showed that galaxies out to 20 $\hmpc$ from the ridge display 
infall velocities of the order of 1000 km s$^{-1}$ (see also Willick 1990;
Eisenstein \etal\ 1997).  Unfortunately, there is no way to disentangle the two
effects from \xip\ alone, and we will not discuss the large-scale
distortions further in this paper. 

Figure~\ref{xipi_morp} shows \xip\ for E--19.5 and S--19.5. 
The visual difference between the two maps is impressive; 
the ellipticals display a huge small--scale elongation of the contours,
while \xip\ for the spirals is much more isotropic. 
This figure demonstrates directly how different the 
dynamical behaviors of the two populations are, and how their
real--space correlation functions are mapped into redshift space in
very different ways.

  We now go on to quantify the real space correlation function, and
the redshift distortions, from the observed \xip. 

\subsection{The Real--Space Correlation Function}
\label{sec:xi(r)}

We project \xip\ onto the $r_p$ axis by integrating over the 
dimension on which the redshift--space distortion acts, giving 
a quantity that is independent of the form and amount of the distortion
itself,
\begin{equation}
w_p(r_p) \equiv 2 \int_0^{\infty} dy\, \xi(r_p,\pi) = 
2 \int_0^{\infty} dy \, \xi\left[(r_p^2 + y^2)^{1/2}\right]\, ,
\label{wp}
\end{equation}
where the second equality follows from the independence of the
integral on the redshift-space distortions.  In the expression on the
right-hand side, $\xi$ is the {\it real--space\/} 
correlation function, evaluated at $r=(r_p^2 + y^2)^{1/2}$.  Modelling
$\xi(r)$ as a power law, $\xi(r) = (r/r_0)^{-\gamma}$ allows us to
carry out the integral analytically, yielding
\begin{equation}
w_p(r_p)=r_p\left({r_0\over r_p}\right)^\gamma {\Gamma({1\over 2})\,
\Gamma({\gamma-1\over 2}) \over \Gamma({\gamma\over 2})}
\end{equation}
where $\Gamma$ is the Gamma function.   
We choose $\pi_{up}$, the upper integration limit in 
Eq.~(\ref{wp}), to be large enough to give a stable estimate of
$w_p$. 
\begin{table*}
\begin{center}
\begin{tabular}{lcc}  \hline\hline
Sample & $r_0\;(\hmpc)$ & $\gamma$ \\
\hline
PP--19  &$5.95_{-0.31}^{+0.27}$ & $1.93_{-0.07}^{+0.04}$\\
PP--19.5&$6.95_{-0.32}^{+0.37}$ & $1.88_{-0.07}^{+0.05}$\\
PP--19.5--NOP&$6.55_{-0.36}^{+0.34}$ & $1.86_{-0.05}^{+0.04}$\\
PP--20  &$7.05_{-0.59}^{+0.47}$ & $1.72_{-0.10}^{+0.09}$\\
S--19   &$4.55_{-0.36}^{+0.36}$ & $1.65_{-0.07}^{+0.06}$\\
S--19.5 &$5.55_{-0.45}^{+0.40}$ & $1.73_{-0.08}^{+0.07}$\\
S--19.5--NOP &$5.05_{-0.48}^{+0.54}$ & $1.76_{-0.10}^{+0.08}$\\
S--20   &$5.05_{-0.65}^{+0.61}$ & $1.85_{-0.09}^{+0.08}$\\
E--19.5 &$8.35_{-0.76}^{+0.75}$ & $2.05_{-0.08}^{+0.10}$\\
\hline
\end{tabular}
\end{center}
\caption{Best--fit parameters of the real--space correlation function from
$w_p(r_p)$.
}
\label{wpfit_tab}
\end{table*}                                                                   
\begin{table*}
\begin{center}
\begin{tabular}{lll}  
\hline\hline
Sample  & $\sigma_{12}(1)$ ($F$=0) &
$\sigma_{12}(1)$ ($F$=1)\\  
\hline
PP--19  & $775^{+85}_{-65}$& $855^{+85}_{-75}$\\
PP--19.5& $735^{+155}_{-115}$ &$805^{+155}_{-115}$\\
PP--19.5--NOP& $625^{+125}_{-85}$ &$725^{+135}_{-95}$\\
PP--20  & $525^{+155}_{-115}$ &$465^{+145}_{-105}$\\
S--19    &$205^{+75}_{-55}$ &$295^{+75}_{-55}$   \\
S--19.5  &$255^{+95}_{-65}$ &$345^{+95}_{-65}$\\
S--19.5--NOP  &$235^{+115}_{-75}$ &$325^{+125}_{-85}$\\
S--20    &$415^{+465}_{-245}$ &$485^{+465}_{-245}$\\
E--19.5  &$815^{+245}_{-165}$ &$865^{+250}_{-165}$\\
\hline
\end{tabular}
\end{center}
\caption{Summary of the best estimates of the pairwise velocity dispersion 
between 0 and 1 h$^{-1}$ Mpc, $\sigma_{12}(1)$, for the two cases $F=0$ 
(free streaming with the Hubble flow), and $F=1$ (stable clustering).  All estimates are in km sec$^{-1}$.} 
\label{sigma_tab}
\end{table*}                                                                   
\hfil

For the PP--19 sample, \wp\ is quite insensitive to $\pi_{up}$ in the
range $ 20\hmpc < \pi_{up} < 30 \hmpc$ for $r_p < 10 \hmpc$.  For
larger values of $r_p$, \wp\ is in fact fairly sensitive to
$\pi_{up}$, but since we are primarily interested in the redshift
distortions on small scales, this has no effect on our result. 

The observed \wp\ and the best--fit power law for the complete samples
are shown in Figure~\ref{wpfit_vlim}, together with likelihood 
contours on $r_0$ and $\gamma$, while the results of the fits are
reported in Table~\ref{wpfit_tab}.   Note how well the power--law model
fits the data\footnote{Note also, however,
that $w_p(r_p)$ is an integral over $\xi(r)$, and therefore small deviations 
{}from the power--law model in the latter function are averaged out in the
former.}.  Error bars are given
by the scatter over 100 bootstrap realizations and the fit is performed
as discussed in \S~\ref{sec:ML}.
There is evidence of a growth of the correlation length 
with increasing sample depth and intrinsic luminosity.  This
is most significant ($\sim 3\sigma$) between PP--19 and PP--19.5;
$r_0$ does not grow significantly between PP--19.5 and PP--20. This
is in qualitative agreement with the results of  
Iovino \etal\ (1993) using a previous version of this sample, 
and the results of Loveday \etal\ (1995) on the APM--Stromlo
redshift survey,  but is in contrast with Hamilton (1988), who found
that the luminosity dependence of $r_0$ was most significant at the
highest luminosities.  Table~\ref{wpfit_tab} also shows 
a similar trend for the spiral--only samples.  Thus 
even within morphological classes, a luminosity
dependence of clustering does exist (Iovino \etal\ 1993; cf., their
Figure~12).  
We have checked the sensitivity of these results to magnitude errors
at the faint end by cutting 
the PP--19, PP--19.5 and PP--20 samples at a corrected magnitude $m_{Zw}=15.2$,
and re--computing \xip, $w_p(r_p)$ and the best fit with a power--law \x. 
This is a fairly conservative selection, reducing each of the three 
subsamples
by $\sim 30\%$ in number (to 882, 740 and 503 galaxies respectively).
For these three samples, we obtain the following estimates for ($r_0$,
$\gamma$):
 ($5.95_{-0.30}^{+0.34}\hmpc$, $1.92_{-0.07}^{+0.06}$), 
($6.85_{-0.41}^{+0.40}\hmpc$, $1.90_{-0.06}^{+0.11}$), and
($7.45_{-0.64}^{+0.64}\hmpc$, $1.74_{-0.10}^{+0.11}$), respectively.  Comparison of these
values with those in Table~\ref{wpfit_tab} shows that the results
are very robust and that our conclusions are unaffected by any magnitude
bias affecting the faint end of the Zwicky catalogue. 

Figure~\ref{wpfit_types} shows one of our principal results,
the relative clustering strength of early-- and late--type
galaxies, as described by the real--space correlation function.
Both the slope and correlation length are 
significantly different in the two samples (Table~\ref{wpfit_tab}).

The scale dependence of the 
relative bias $b_{ES}$ of early to late--type galaxies is then simply:
\begin{equation}
b_{ES}(r) = \left({\xi_E(r) \over \xi_S(r)}\right)^{1\over 2} = b_1\cdot
r^{-\nu} \,\,\, , 
\end{equation}
where $b_1$ is the value at $1\hmpc$ and $\nu = (\gamma_S - \gamma_E)^{0.5}$. 
We find $b_{ES}(r)= (2.0 \pm 0.4)
r^{-0.16\pm 0.08}$, where the error bars have been computed
using standard error propagation.  At $r = 5\hmpc$, we find $b_{ES} =
1.6 \pm 0.4$.  Hermit \etal\ (1996) compute a similar relative bias
factor from the Optical Redshift Survey (Santiago \etal\ 1995), but in
redshift space,  finding an average value $\sim 1.5$ between 1 and $10 \hmpc$.
Their analysis does not take into account the differences in
redshift--space distortions between the two classes of galaxies that
we have stressed here.  
Loveday \etal\ (1995) use both the APM catalogue and the 
sparsely--sampled subsets of galaxies that form the APM--Stromlo 
redshift survey; inverting
the angular correlation function $w(\theta)$ for two subsamples
limited to $b_J=16.57$ 
they find $r_0 = 7.76\pm 0.35 \hmpc$, $\gamma=1.87\pm 0.07$
for early-type galaxies and $r_0 = 4.49\pm 0.13 \hmpc$, $\gamma=1.72\pm 0.05$
for late-type galaxies.  This is in good agreement with our direct
estimates from the spatial function.  However, the APM--Stromlo
data are too sparse for being able to compute \xip\ directly 
{}from the morphological subsamples,
so that an estimate of the pairwise velocity dispersion 
$\sigma_{12}(r)$ cannot be obtained.  

The correlation length we find for all spiral galaxies 
is significantly larger than that of \iras\ galaxies in the 1.2 
Jy redshift survey, $r_0 ({\rm 1.2\, Jy}) = 3.76^{+0.20}_{-0.23} 
\hmpc$, while the slope $\gamma({\rm 1.2\, Jy}) = 
1.66^{+0.12}_{-0.09}$ is similar.  This has the interesting
implication that the  
relative bias of spiral galaxies to \iras--selected galaxies is 
independent of scale, at least below 10 \hmpc.   Since \iras\ galaxies tend
to be of type Sb and later, we have defined a volume-limited sample
to $M = -19$, containing 321 galaxies with types between Sb
and Irr.  In fact, for these we find a lower correlation length, 
$r_0 = 4.05^{+0.57}_{-0.75} \hmpc$, and a similar logarithmic slope,
$\gamma = 1.55^{+0.11}_{-0.13}$, in excellent agreement with \iras\
galaxies.

\subsection{The Pairwise Velocity Dispersion}
\label{sec:sigma-12}

The quantity \xip\ can be expressed as an integral over the product of
the real-space correlation function $\xi(r)$, and the distribution
function of the line--of--sight components $w_3$ of relative
velocities for pairs with separation $r$, $f(w_3|r)$ (\cite{f94b})
%
%
If $y$ is the component of $r$ along the line of sight,
then $w_3 = H_0(\pi - y)$ and the integral can be 
written as (\cite{peebles80}, \cite{f94b})
\begin{equation}
1+\xi(r_p,\pi) = H_0 \int_{-\infty}^{+\infty} dy \left\{1+\xi\left[
(r_p^2 + y^2)^{1\over 2} \right]\right\}\,f\left[H_0 (\pi - y)|r\right]
\,\,\, .
\label{conv1}
\end{equation}
This expression gives a {\it description} of the effect of a peculiar 
velocity field on $\xi(r)$, but does not represent a self--consistent
dynamical treatment of the density and velocity fields, which are
clearly interdependent (\cite{f95}).  We do not have any {\it a priori}
information, therefore, on the functional form of the 
distribution function $f$.  Peebles (1976)
first showed that an exponential distribution best fits the observed data,
a result subsequently confirmed by $N$--body models (e.g., Zurek \etal\ 1994).
With such a choice, Eq.~(\ref{conv1}) becomes 
\begin{equation}
1+\xi(r_p,\pi) = H_0 \int_{-\infty}^{+\infty} dy \left[1+\xi(r)
\right]\,{1\over \sqrt{2} \sigma_{12}(r)}\,{\rm exp}\left\{ -\sqrt{2}H_0
 \left|{\pi - y\left[1+{v_{12}(r) \over H_0 r}\right] \over \sigma_{12}(r)}\right| \right\}
 \,\,\, ,
\label{xp_model}
\end{equation} 
\noindent where $r^2 = r_p^2 + y^2$, $v_{12}(r)$ is the mean relative velocity
of galaxy pairs with separation $r$, and $\sigma_{12}(r)$ is the pairwise
one--dimensional velocity dispersion along the line of sight. 

\cite{f94b} show that it is very difficult to model the dependence of $v_{12}$ on the 
separation $r$.  This is made particularly difficult in our case; 
our sample covers too small a volume to allow a determination of the
large-scale streaming from the compression of the contours of 
\xip\ (cf., \S~\ref{xp-shape}).
For this reason, we do not follow \cite{f94b} in a detailed analysis of
the mean streaming, and instead, we limit ourselves to the simple 
streaming model introduced by \cite{dp83}, based on the 
similarity solution of the BBGKY equations: 
\begin{equation}
v_{12}(r) = -H_0 r {F \over 1+\left({r\over r_0}\right)^2} \,\,\, .
\label{eq:v12}
\end{equation}
We wish to fit Eq.~(\ref{xp_model}) to the observed \xip\ to constrain
$\sigma_{12}(r)$.  We are interested in particular in $\sigma_{12}(1)$, 
the pairwise
velocity dispersion for scales smaller than 1 \hmpc, and thus we carry
out all fits to the quantity 
\begin{equation} 
\xi(\pi) = \int_0^1 d r_p\, \xi(r_p,\pi)\qquad,
\label{eq:xipi} 
\end{equation}
following \cite{f94b}\footnote{Our definition of $\xi(\pi)$ differs
from that of \cite{f94b} by an unimportant normalization factor -- cf., their
Eq.~(7).}. In
practice, because we have calculated \xip\ in 1 \hmpc\ bins,
$\xi(\pi)$ is simply the value of \xip\ in the first bin of $r_p$. 
We assume further that $\sigma_{12}(r)$ is a weak function of
separation $r$ (\cite{dp83}),
so that it can be treated as a single free parameter, $\sigma_{12}(1)$.  
Figure~\ref{xpfit_types_01} shows the results of two--parameter fits 
of the model of
Eqs.~(\ref{xp_model}) and (\ref{eq:v12}) to $\xi(\pi)$ for the E--19.5
and S--19.5 subsamples. We use the best-fit values
of $r_0$ and $\gamma$ from Table~\ref{wpfit_tab} appropriate for each
subsample; errors and covariances of $\xi(\pi)$ are calculated
consistently, as described in \S~\ref{sec:ML}.  The quantity $F$ is
very poorly constrained by these data (Figure~\ref{xpfit_types_01}): the
free-streaming on these small scales is quite small. 

We therefore estimate \sig1\ for the two cases $F=0$ (free expansion 
of pairs with the Hubble flow), and $F=1$ (stable clustering).
This second case is most probably the one closer to a realistic model: 
Jain (1996) shows that the stable clustering hypothesis ($F=1$) should be
a good approximation at the present epoch for scales of the order of,
or smaller than $0.7 \hmpc$.  
We thus use the values with $F = 1$ in our discussion below.

The value of \sig1\ is of order 800 \kms\ for PP--19 
and PP--19.5, but drops below $500 \kms$ for PP--20, consistent with the 
more isotropic contours of \xip\ for this case
(Figure~\ref{xipi_second}).  We interpret this as due to the smaller
effect rich clusters, and in particular the Perseus cluster, have on 
the larger volume of PP--20, as we shall see in the next section.  
Notice the very significant
factor of two difference between the \sig1\ for early-- and late--type
galaxies at $M_{Zw} = -19.5$, a dramatic indication of the effect of
cluster cores on the determination of \sig1.  We now turn to a direct
demonstration of the sensitivity of \sig1\ to the presence of rich
clusters in the sample.

\subsection{Stability of $\sigma_{12}$ for Late--Type Galaxies}
\label{sec:noperseus}

Marzke \etal\ (1995) have discussed in detail the effect of the contribution
of cluster galaxies to the small--scale pairwise velocity dispersion.
The pairwise velocity dispersion is a pair-weighted statistic, and
thus it is heavily weighted in regions of high density, i.e.,
clusters.  Because galaxies in clusters have an intrinsically high velocity
dispersion, the inclusion or exclusion of clusters can have a dramatic
effect on $\sigma_{12}$.  
Marzke \etal\ showed that the estimates of $\sigma_{12}(r)$ fluctuate from one sample 
to the other due to the significant variations in the number of clusters
even over volumes as large as those of the CfA2 and SSRS2 surveys.
Guzzo \etal\ (1996) showed that \sig1\ dropped from $\sim 800 \kms$
to $\sim 600 \kms$ in PP--19, after removing the Perseus cluster.
Thus the removal of a single dominant cluster can
significantly affect the pairwise velocity dispersion. 

Here we explore further the stability of $\sigma_{12}(1)$,
in the case of spiral galaxies.   Using spiral--only samples, we are
in practice filtering out the high--density non--linear regions that
would otherwise get such high weight in $\sigma_{12}(1)$.
The result of excluding the Perseus cluster from the PP--19.5 sample
is visually shown by the changes in \xip\, in the two top panels of 
Figure~\ref{xi4_spir}.
The differences between the contours in the two panels [and the  
corresponding values of $\sigma_{12}(1)$, reported in Table~\ref{sigma_tab}],
can be compared to those produced by the same operation on the S--19.5 sample
(bottom).  While the effect on PP--19.5 is relevant [although less
dramatic than it was found for PP--19 by Guzzo \etal\ (1996) for PP--19,
due to the larger volume and the consequent reduced weight of the Perseus 
cluster], the two bottom panels of Figure~\ref{xi4_spir} are virtually 
identical, and so are the estimated \sig1\ .

Table~\ref{sigma_tab} also gives $\sigma_{12}(1)$ for spiral samples
limited to $M_{Zw} = -19$ and $-20$.  
Table~\ref{sigma_tab} indicates that \sig1\ for spirals 
lies consistently between 300 and 350 \kms.  This agrees with the
\cite{f94b} value for \iras\ galaxies  $317^{+40}_{-49} \kms$, the
\cite{marzke} value for 
galaxies outside of $R \ge 1$ Abell clusters $295\pm 99 \kms$,
and the original determination by \cite{dp83}, which undersampled the Virgo
cluster in CfA1 (Somerville, Davis \& Primack 1997).

It is interesting to discuss the similarity of \sig1\ for  spiral and 
\iras\ galaxies, in the light of their different correlation lengths 
($r_0 \simeq 5.5 \hmpc$ and $r_0 \simeq 3.8 \hmpc$, respectively).  \iras\
galaxies are mostly late--type spirals, and indeed we showed above that
if we compute $\xi(r)$ for this subclass, we recover $r_0 \simeq 4.0 
\hmpc$, in agreement with \iras\ galaxies (cf., Giovanelli,
Haynes \& Chincarini 1986, and Iovino \etal\ 1993, who showed
that there is a continuity in the clustering strength within the spiral
class, with Sc's being less clustered than Sa's). 
The similar value of 
\sig1, on the other hand, may simply indicate that the dynamics of
\iras\  galaxies and all spirals are governed by the same fluctuations
in the underlying matter density field.   Indeed, \sig1\ for galaxies
of type Sb and later is found to be $255^{+105}_{-75} \kms$, in
statistical agreement with the spiral sample as a whole.


\section{SUMMARY AND DISCUSSION}
\label{sec:summary}

The main conclusions we have reached in this paper can be summarized as
follows.

\begin{itemize}

\item We see very strong small--scale redshift-space distortions in
the Perseus--Pisces 
redshift survey.  The distortions are much stronger for early--type
galaxies, as one would expect from the segregation of morphological
types. 

\item We confirm a mild luminosity dependence of clustering for
absolute magnitudes around the knee of the luminosity function, 
$M_{Zw}\sim -19.5$: the correlation length increases from 
$5.95_{-0.31}^{+0.27} \hmpc$ for $M_{Zw}\le -19$ to
$7.05_{-0.59}^{+0.47}\hmpc$ 
for $M\le -20$.  These values are somewhat higher than the 
``canonical'' value of the correlation length based on the \cite{dp83}
analysis of the CfA1 survey, $r_0 = 5.4 \hmpc$.
This latter value is confirmed by the analysis of the ESO Slice
Project (ESP) survey (Bartlett \etal\ 1997), that yields $r_0 = 
4.5^{+0.15}_{-0.17} \hmpc$, while the Las Campanas survey
(Lin 1995), gives $r_0 = 5.00\pm0.14 \hmpc$.
There are probably two reasons for the higher values of $r_0$
measured here.   First, the ``standard'' values quoted above are
estimates of $\xi(r)$ performed on apparent--magnitude--limited
samples.  If there is, as we have seen, a mild luminosity dependence
of clustering, samples that are volume--limited at relatively bright
absolute magnitudes will systematically
measure a higher clustering signal.   Second, the Perseus--Pisces
area is rather rich in clusters of galaxies, so that it probably
over--emphasizes the cluster contribution to $\xi(r)$.  For
comparison, the richest cluster in the CfA1 volume is the Virgo
cluster.
The CfA2+SSRS2 sample
(Marzke \etal\ 1995) covers part of the PP area
and thus has a higher contribution of clusters, resulting in
$ r_0 = 5.97 \pm 0.15 \hmpc$.

\item A meaningful comparison of the relative clustering strength 
of spirals and ellipticals 
can be performed only in real space, i.e. after correcting for the
effect of 
differential redshift space distortions.  A power--law shape, $\xi(r) = 
(r/r_0)^{-\gamma}$, is a good representation of the real--space 
correlation function between 1 and $10\hmpc$ for both ellipticals and 
spirals.  Our best--fit estimate of the power--law parameters gives 
$r_0 = 8.35_{-0.76}^{+0.75} \hmpc$, $\gamma = 2.05_{-0.08}^{+0.10}$
for ellipticals, and $r_0 = 5.55_{-0.45}^{+0.40} \hmpc$, $\gamma = 
1.73_{-0.08}^{+0.07}$ for spirals.  We model the relative bias 
of the two types of galaxies as a power law with a mild dependence on scale, 
$b_{ES}(r) = (2.0 \pm 0.4) \left({r /1 \hmpc}\right)^{-0.16 \pm 0.08}$.
Furthermore, we confirm the continuous variation of clustering strength also
within the spiral class.  For late--type spirals (Sb and later),
and irregulars, we estimate $r_0 = 4.05^{+0.57}_{-0.75} \hmpc$, and
$\gamma = 1.55^{+0.11}_{-0.13}$, virtually the same correlation
function as \iras\ galaxies.

\item The quantity \sig1, the measured pairwise velocity dispersion
between 0 and $1 \hmpc$, varies considerably between samples of
different volumes, going from $855^{+85}_{-75}\kms$ of PP--19, 
to $465^{+145}_{-105}\kms$ for PP--20.  This variation is consistent 
with the smallness of the volume sampled.

\item The difference in the measured \sig1\ between early-- and late--type
galaxies is remarkable.  We estimate $\sigma_{12}(1)=865^{+250}_{-165}\kms$
for ellipticals and $\sigma_{12}(1)=345^{+95}_{-65} \kms$ for spirals.  
Contrary to results for the combined sample, the value of 
\sig1\ for spiral galaxies alone is stable to both changes in the sample
volume and the presence of rich clusters.  The consistency of this value
with those measured for non--cluster galaxies (Marzke \etal\ 1995) and
\iras\ galaxies (\cite{f94b}), and the stability among spiral subclasses
(for Sb and later types we measure $255^{+105}_{-75} \kms$), suggests 
that a value \sig1\ in the range 300 -- $350 \kms$ is a good estimate 
of the ``temperature'' of the galaxy flow outside of virialized structures.

\end{itemize}

\acknowledgments

LG gratefully thanks J.P. Ostriker for hospitality at Princeton 
University, where part of this work was done, and J. Bartlett, 
M. Davis and P.J.E. Peebles for fruitful discussions.  We thank the
referee, Ron Marzke, for his valuable comments and suggestions. MAS
acknowledges the generous support of the Alfred P. Sloan Foundation. 
This work was partly supported by grants AST90--23450 and
AST95-28860 to MPH and AST94--20505 to RG.


\clearpage

\figcaption[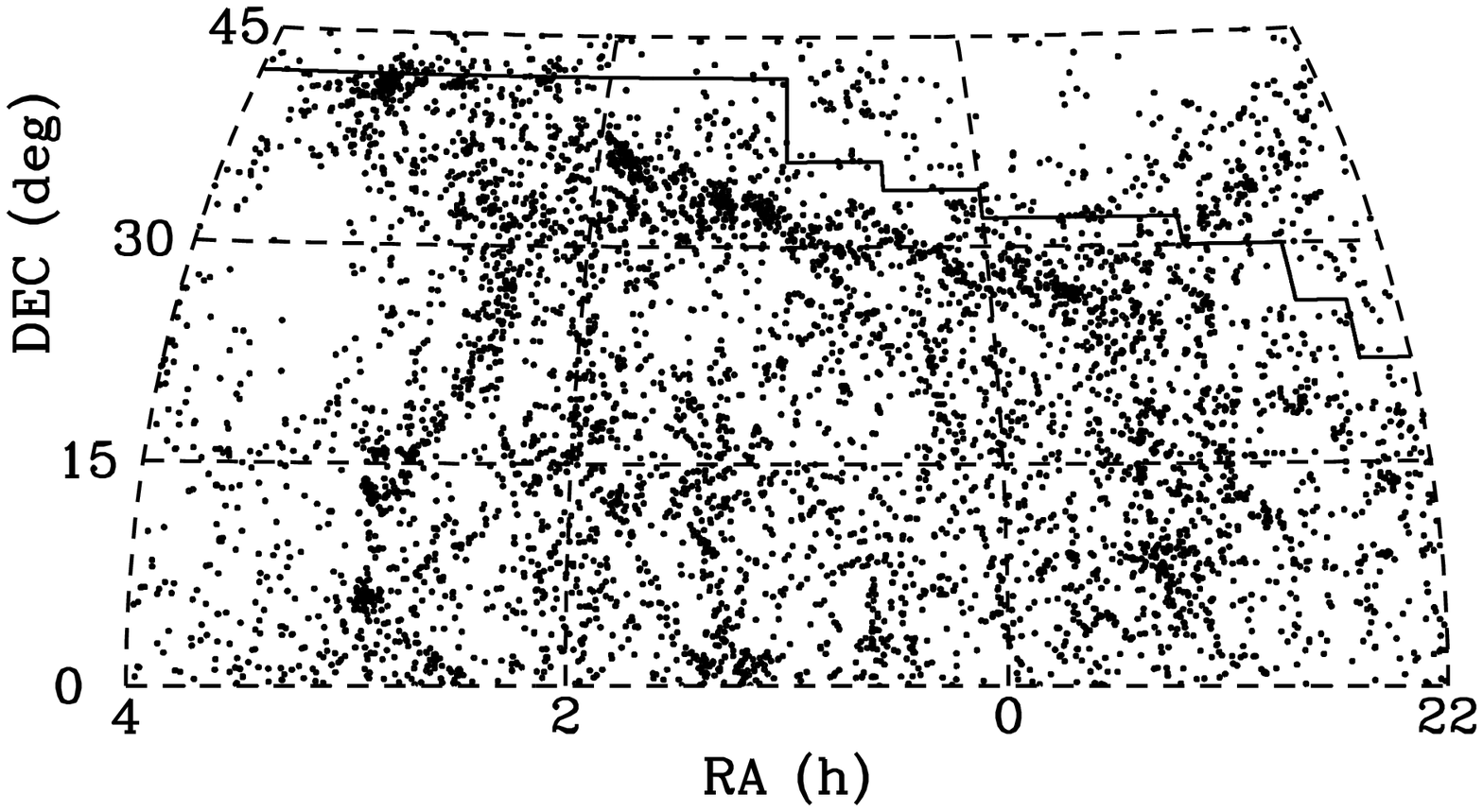]{The distribution on the sky of all galaxies 
with $M_{Zwicky}\le 15.5$
after correction for extinction.  The upper solid line marks the border of the
high--extinction region excluded from the sample.  The large lump of objects
near $\alpha \sim 3.2^h$, $\delta \sim 41^\circ$ is the Perseus cluster. 
\label{survey_map}} 

\figcaption[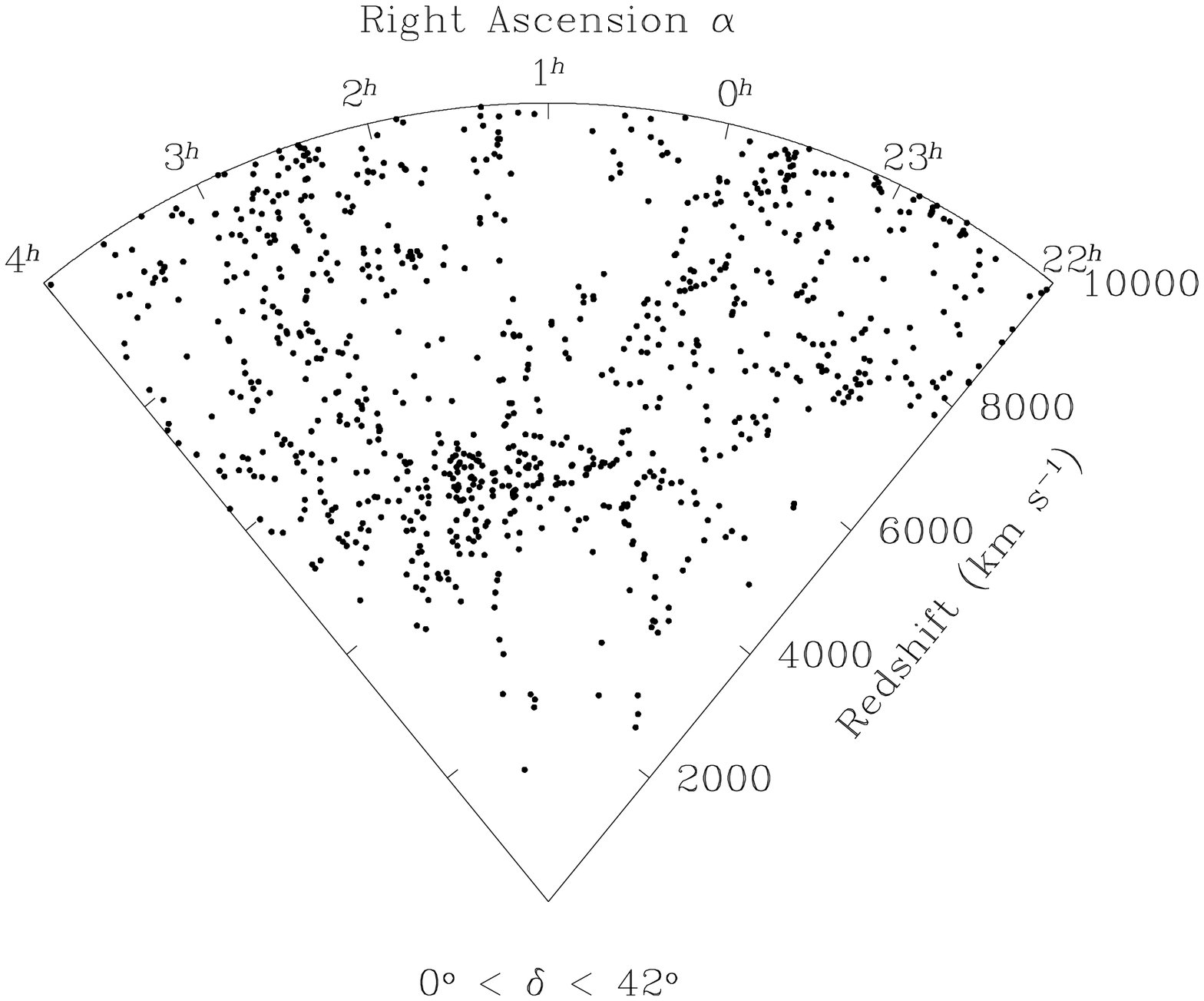, 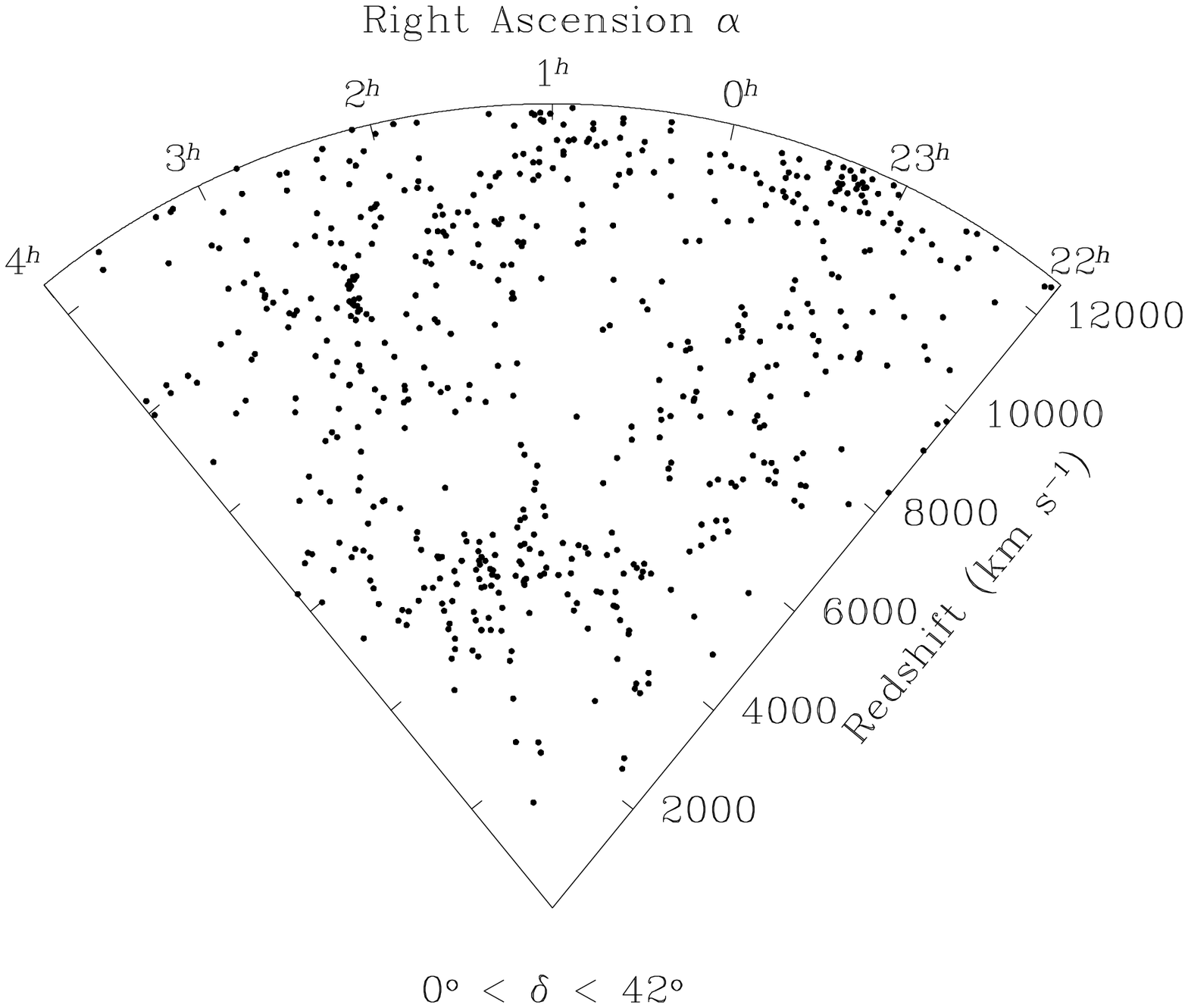]{Galaxy distribution in the 
volume--limited samples to $M_{Zw} = -19.5$ and $M_{Zw} = -20$, 
including all morphological types. 
\label{cone_all}} 

\figcaption[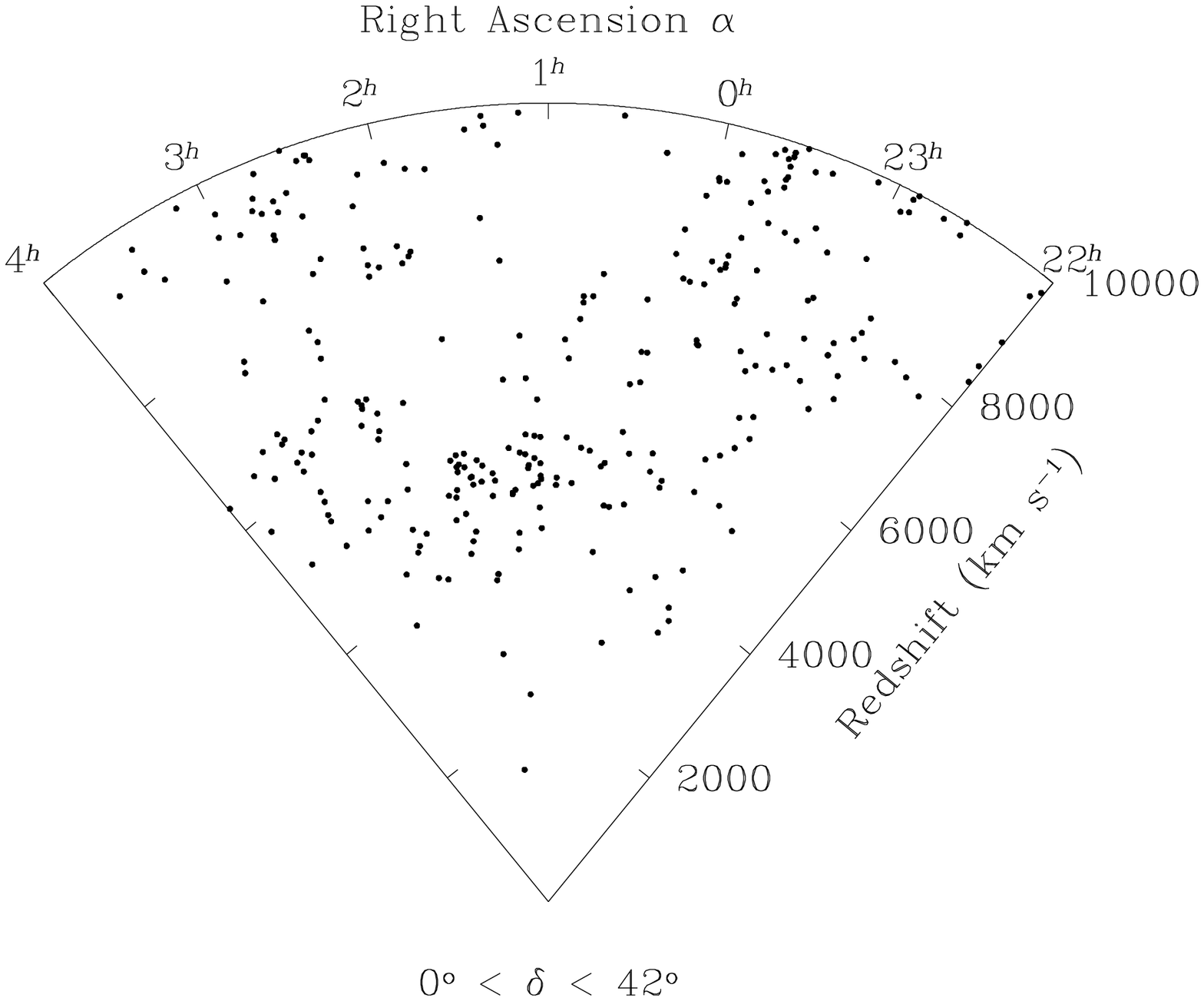, 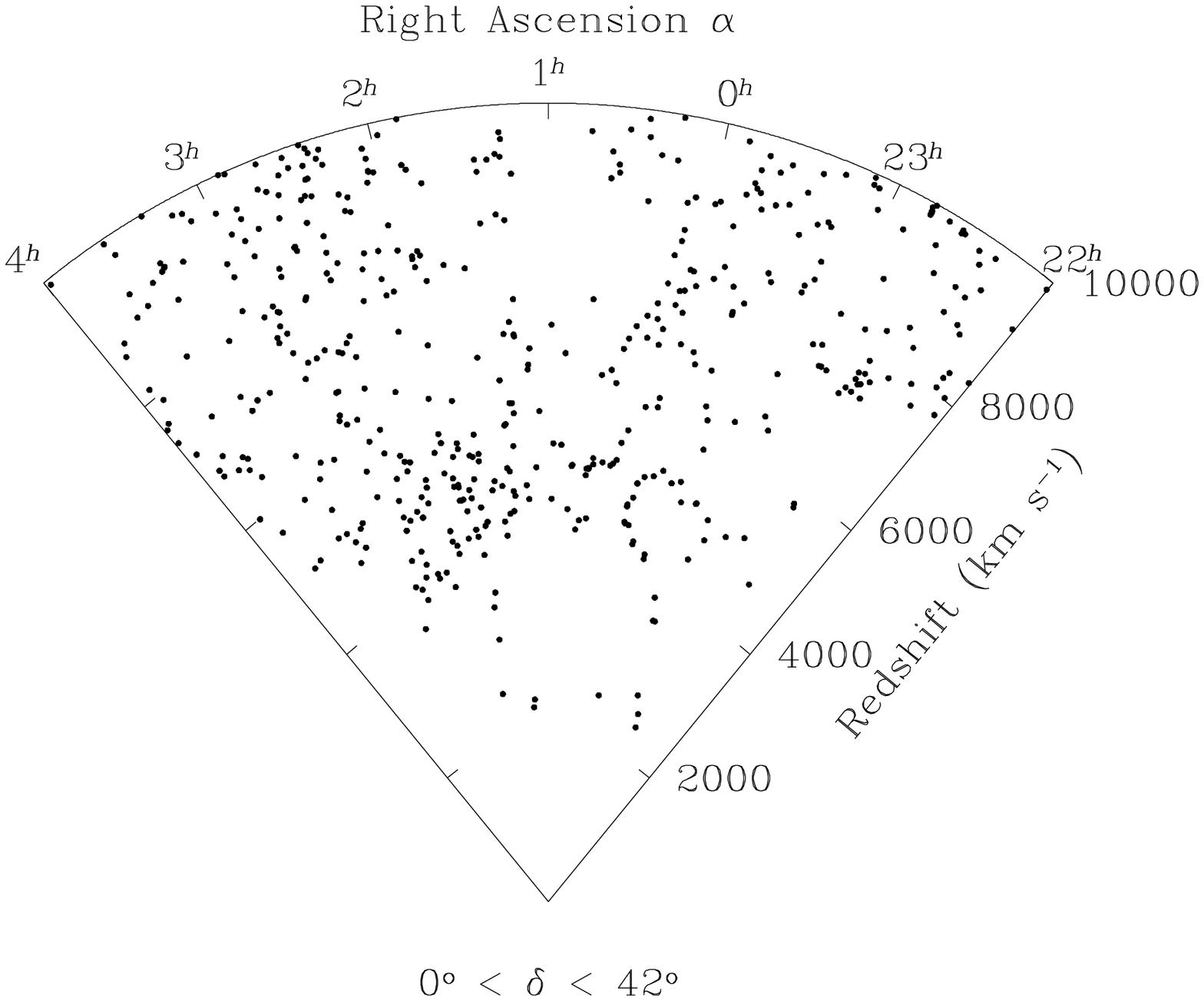]{The spatial distribution 
of early--type and late--type galaxies, volume-limited to $M_{Zw} = -19.5$.
\label{cone_types}}

\figcaption[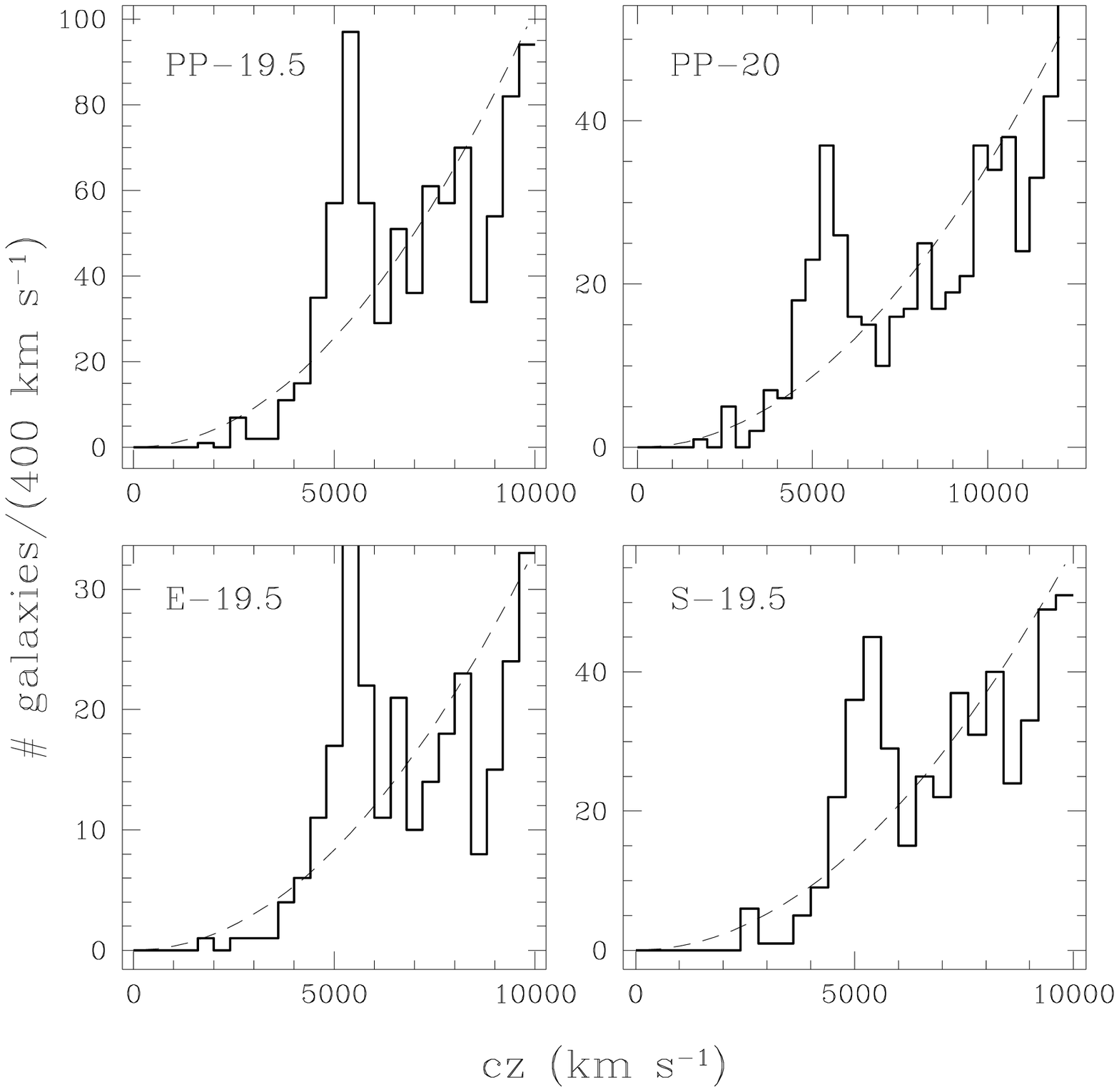]{Histograms of the redshift distribution in four 
representative volume--limited samples.  Top panels: all galaxies.  
Bottom panels: early--type and late--type galaxies separately.  The
dashed lines are the distributions expected in the absence of structure. 
\label{four_hist}} 

\figcaption[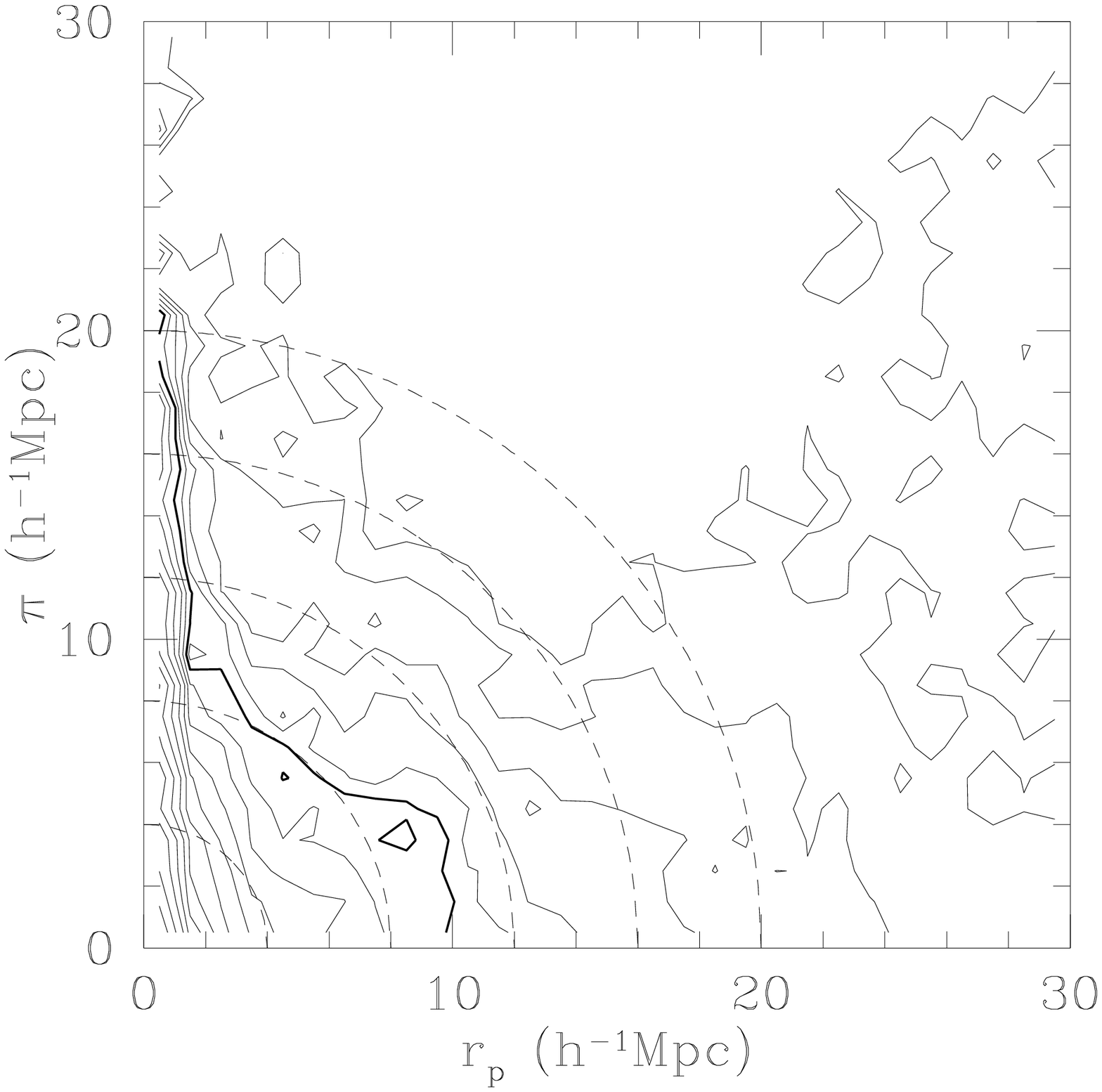,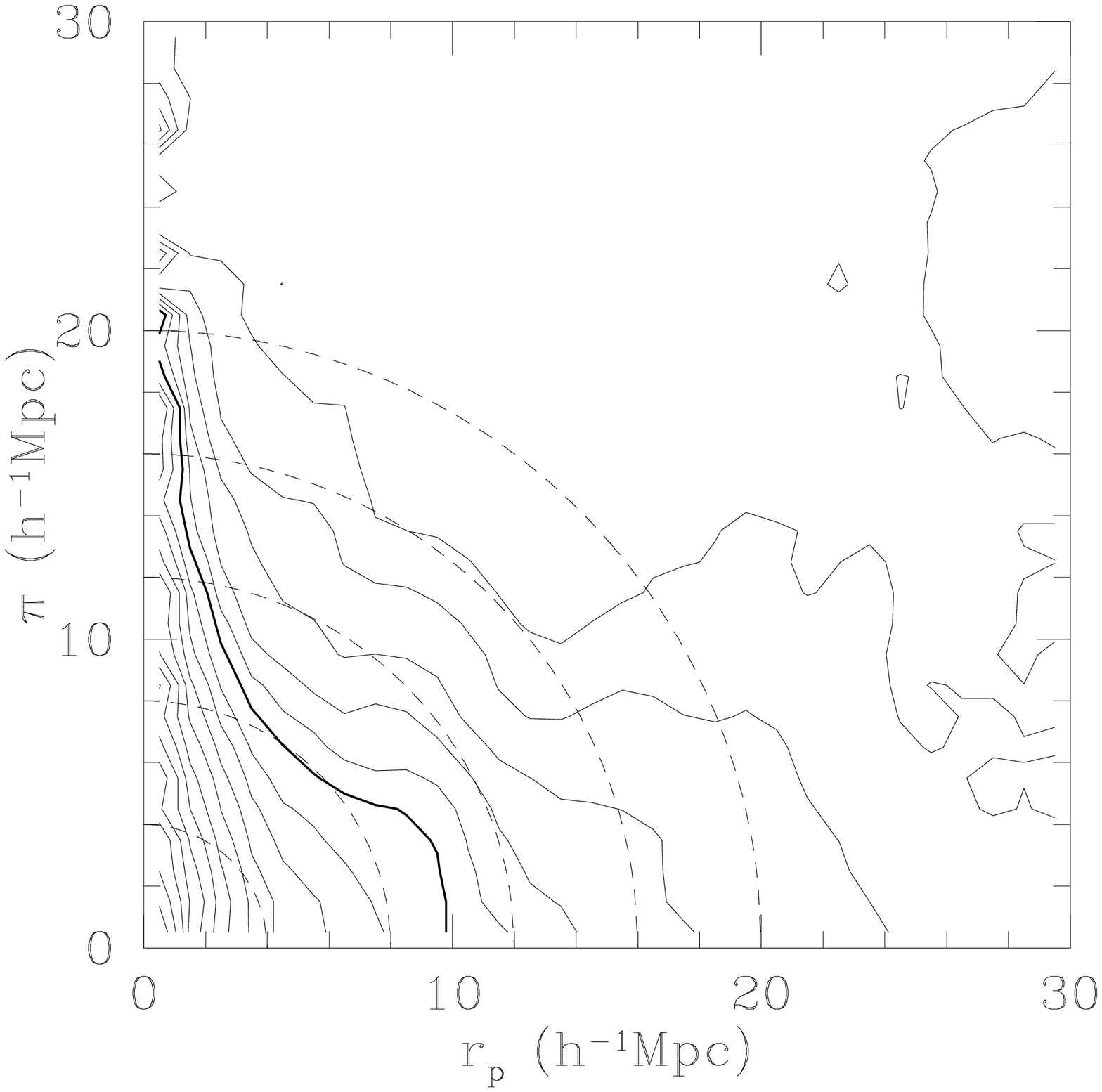]{\xip\ for PP--19. In this and the 
following \xip\ maps, the heavy contour corresponds to $\xi=1$; 
for larger values of $\xi$, contours are logarithmically spaced, 
with $\Delta \log_{10} \xi = 0.1$; below $\xi=1$, they are linearly spaced 
with $\Delta \xi=0.2$ down to $\xi=0$.  The dashed contours represent
the isotropic correlations expected in the absence of peculiar
velocities.  The right--hand panel has been Gaussian-smoothed with an
isotropic filter of width 3~\hmpc.  \label{xipi_first}}

\figcaption[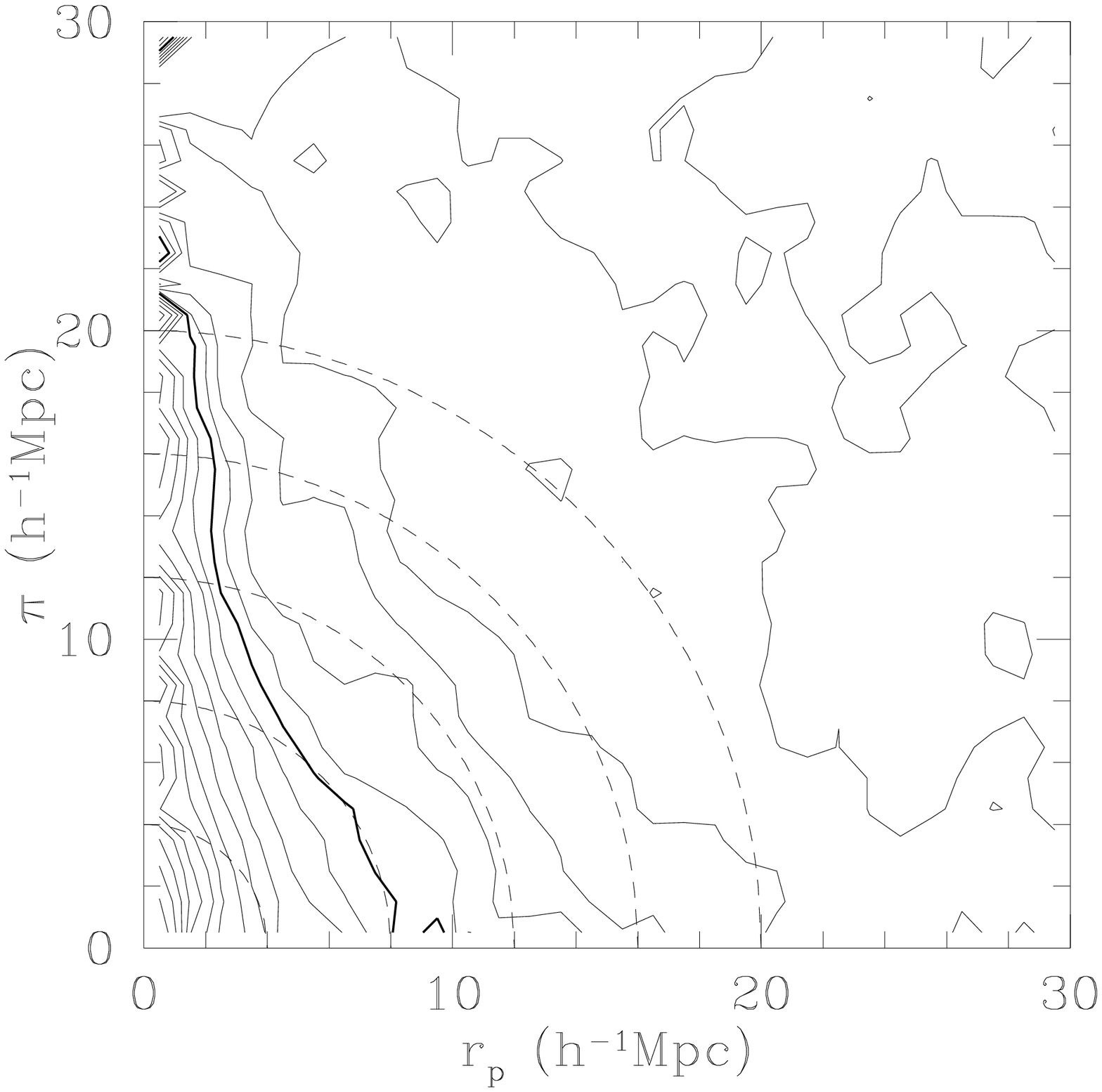,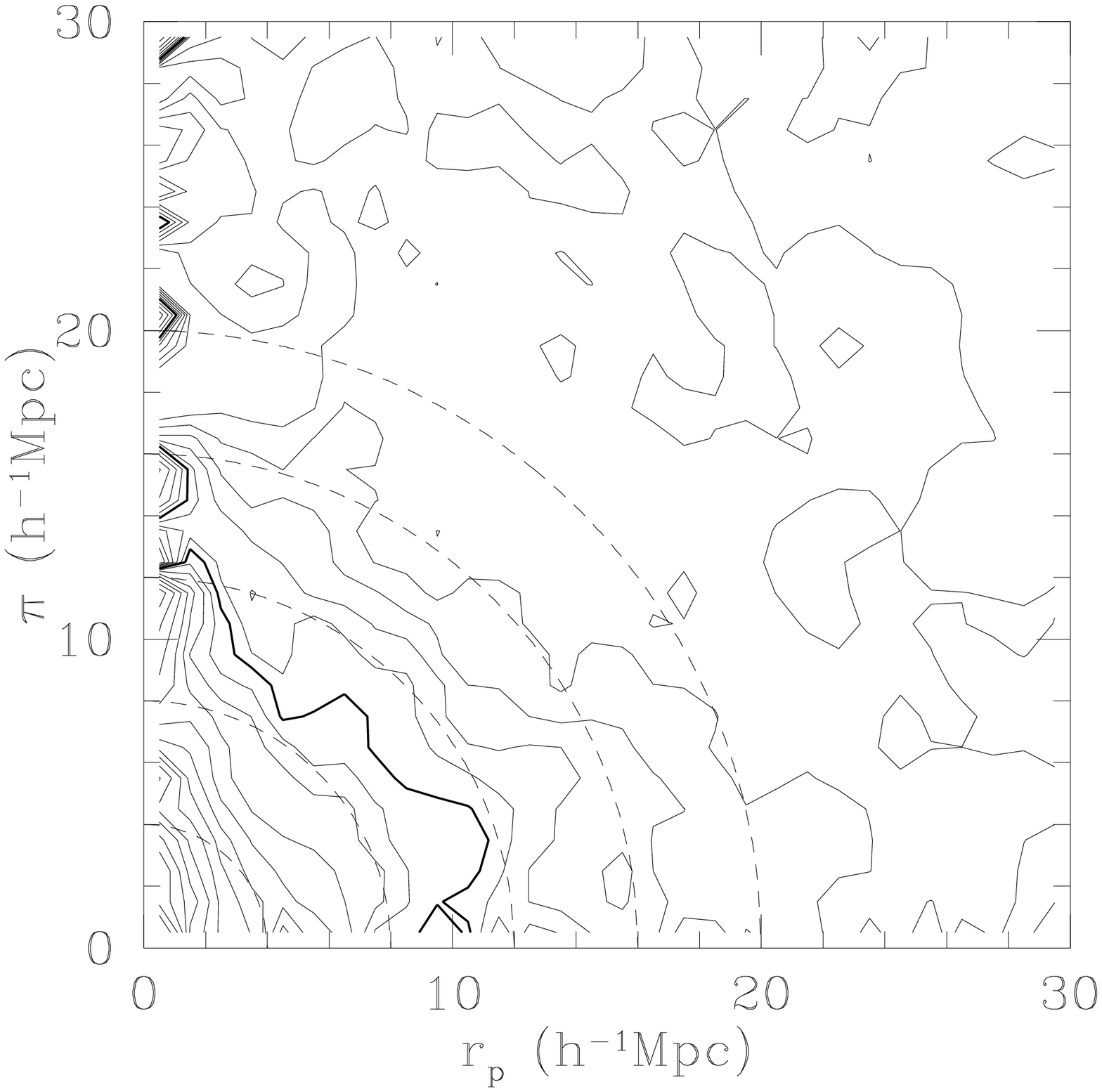]{\xip\ for PP--19.5 and 
PP--20, Gaussian smoothed.\label{xipi_second}}

\figcaption[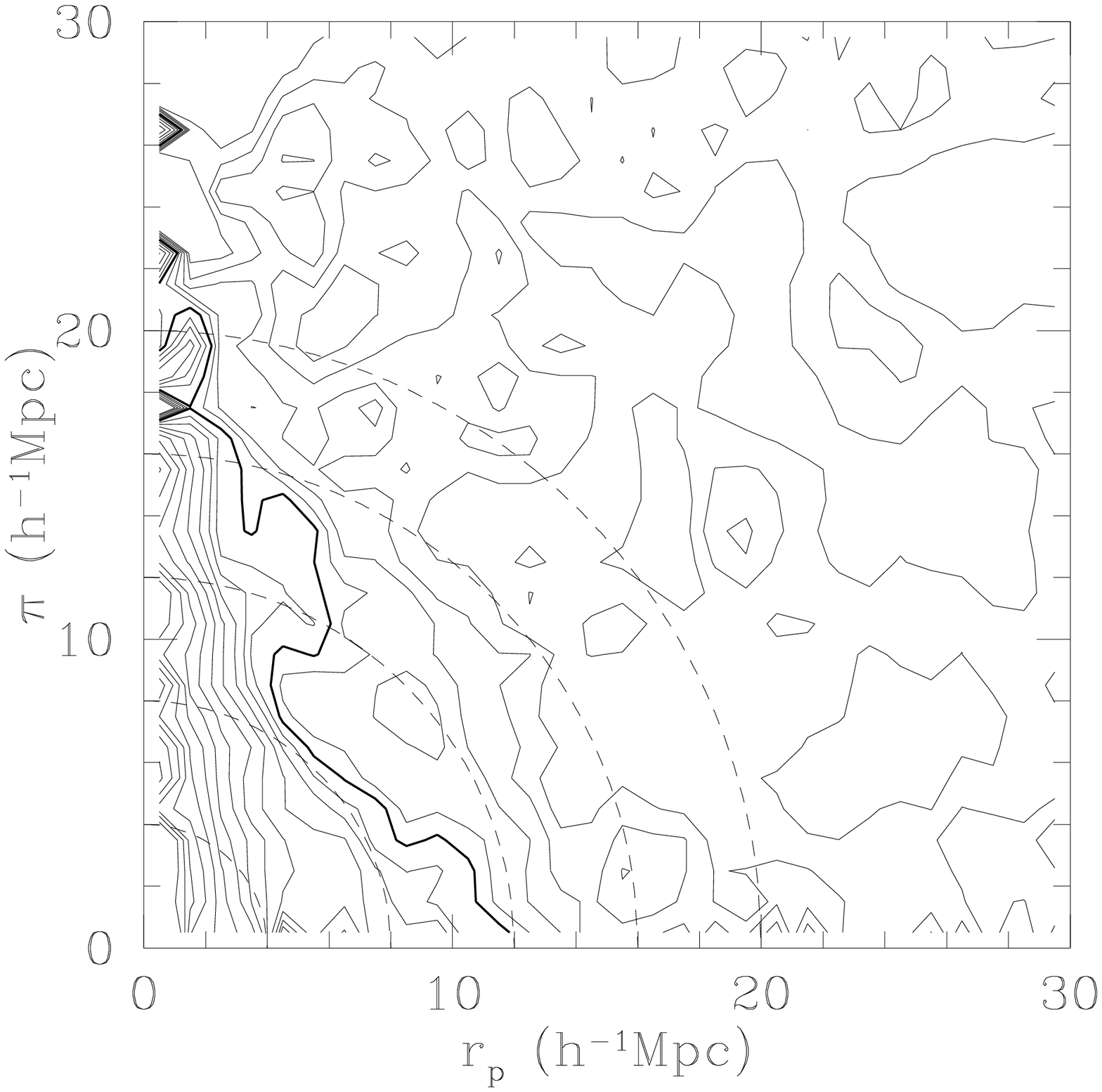,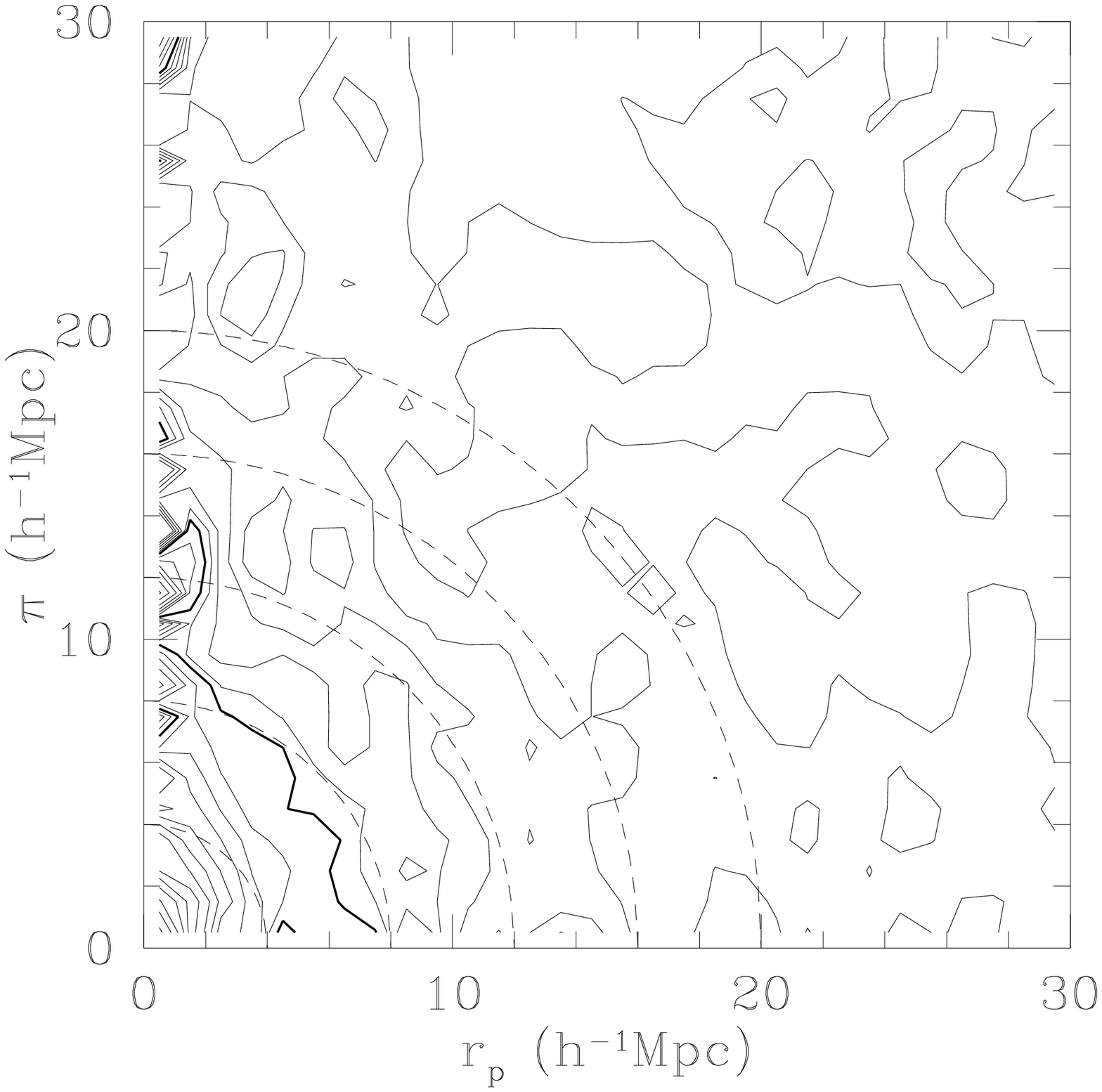]{\xip\ for the early--type 
and late--type samples, E--19.5 and S--19.5, respectively, Gaussian-smoothed.
\label{xipi_morp}}

\figcaption[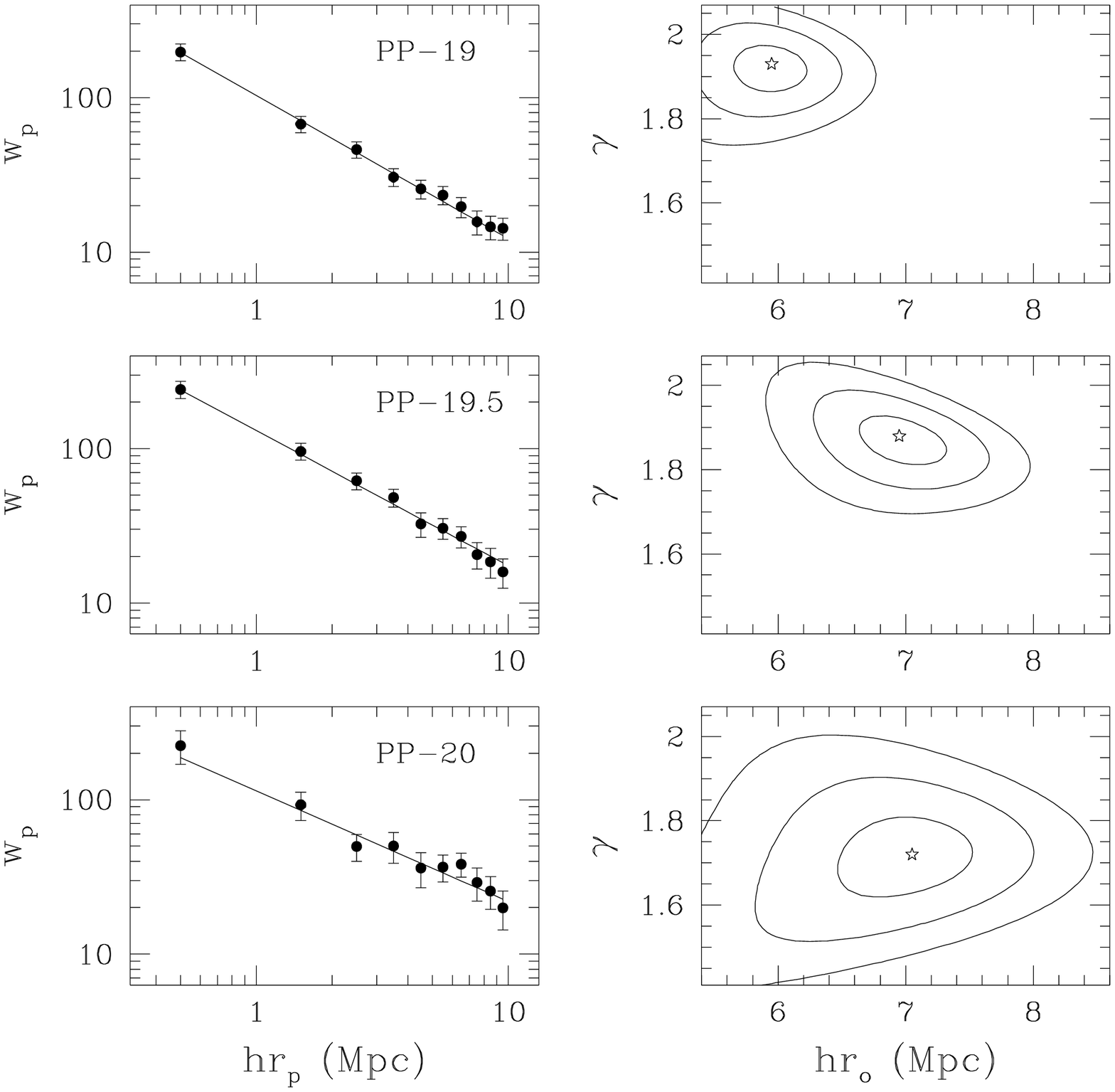]{The projected correlation function $w_p(r_p)$ 
and results of fits of the power--law model, for the three volume--limited 
samples.  The error bars are given by bootstrap resampling. The contours 
give the 68.3\%, 95.4\% and 99.73\% confidence levels on the two parameters 
taken separately. \label{wpfit_vlim}}

\figcaption[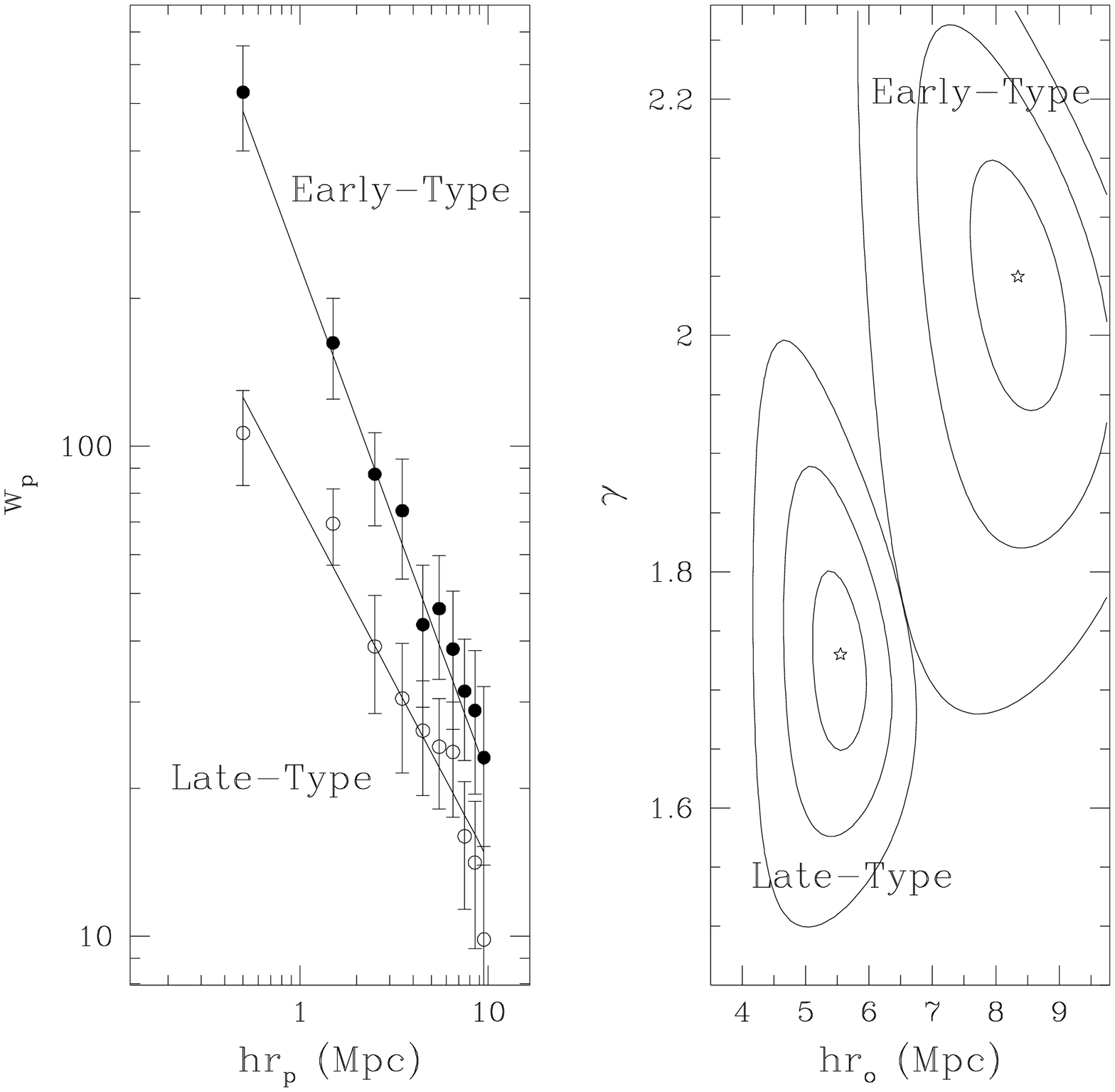]{As in Figure~\ref{wpfit_vlim}, 
for early-- and late--type galaxies, separately.  Note the very significant 
separation in parameter space between the two classes.\label{wpfit_types}}

\figcaption[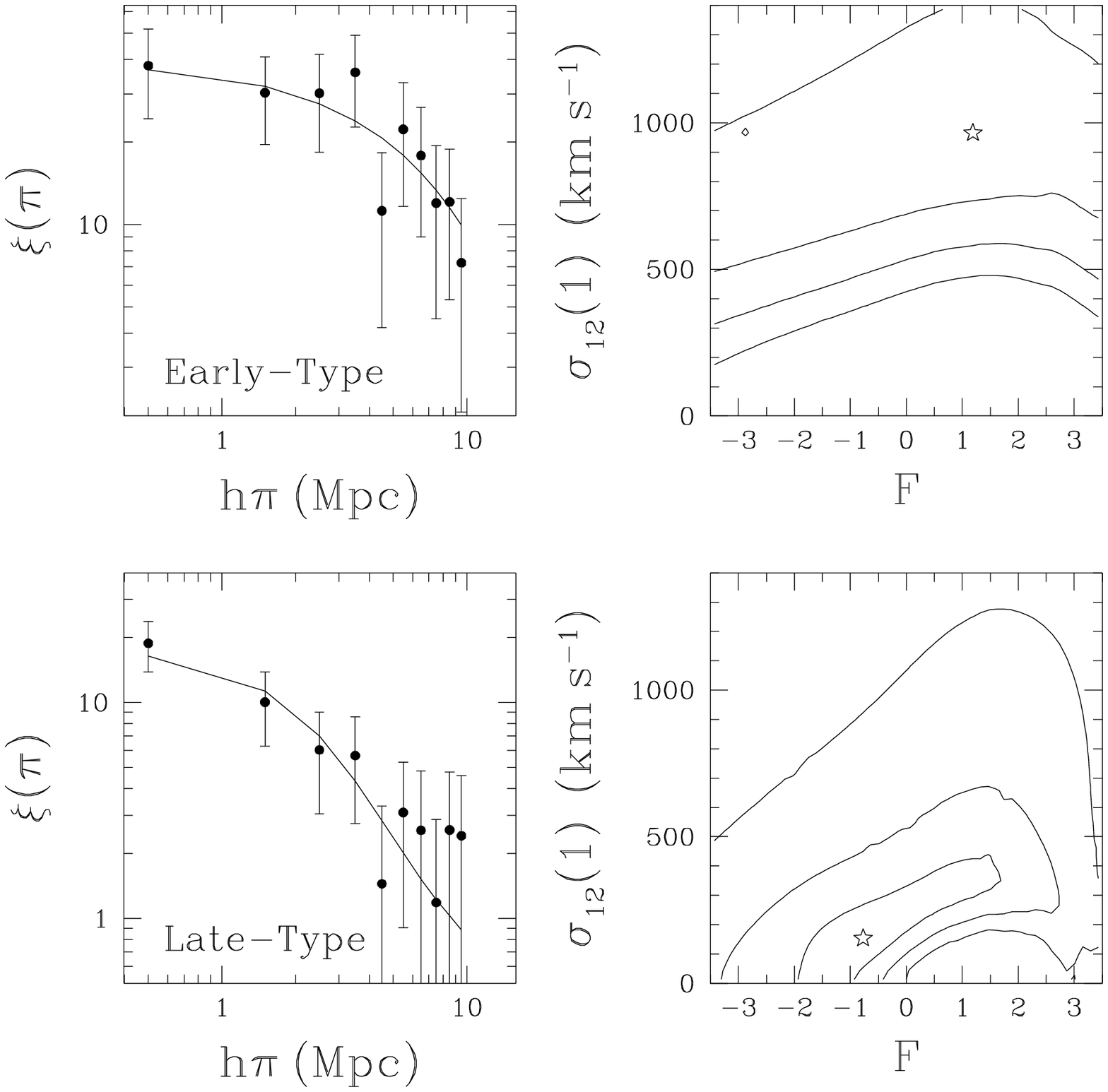]{Example of full two--parameter fits to \xip\ 
based on the model of Eq.~(\ref{xp_model}), to show how poorly constrained
the streaming amplitude $F$ is. Contour levels are as in 
Figure~\ref{wpfit_vlim}. \label{xpfit_types_01}}

\figcaption[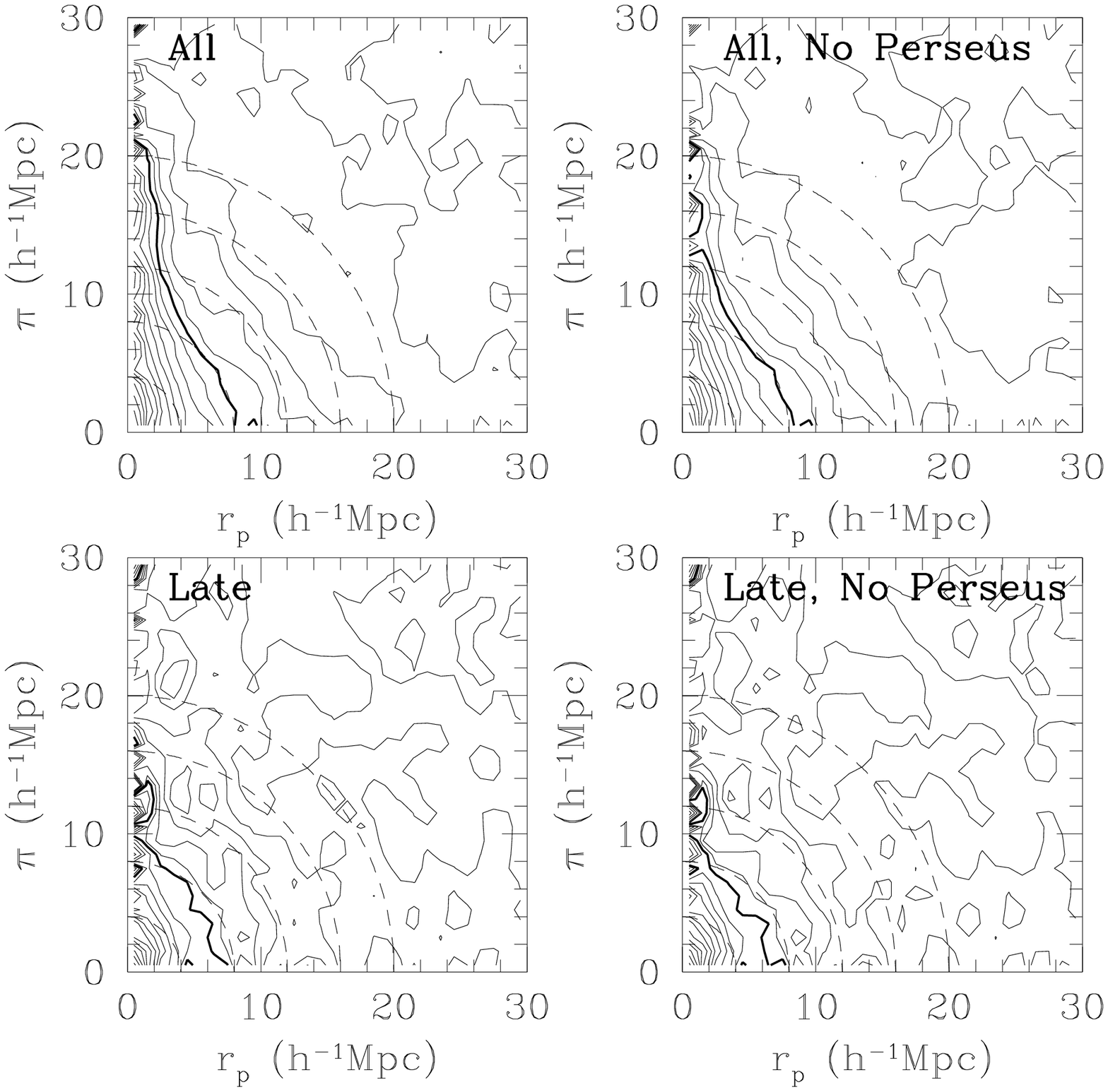]{The effect on \xip\ of removing the largest 
``Finger of God'' from the sample.  Right panels show the effect of excluding
the high--velocity--dispersion Perseus cluster.  Top panels are for
PP--19.5, bottom for S--19.5.  Note the change between the
two top panels, while for spirals the removal of the cluster has
very little effect on the measured \xip.  \label{xi4_spir}}

\end{document}